\documentclass[aps,english,10pt,superscriptaddress,twocolumn,float,floatfix]{revtex4}  
\usepackage{amsmath}
\usepackage{amsthm}
\usepackage{amssymb}
\usepackage{amsfonts}
\usepackage{bbm}
\usepackage{epsfig}
\usepackage{times}
\usepackage{babel}
\usepackage{color}
\usepackage{hyperref}
\usepackage{framed}
\usepackage{changes}

\def\tr{{\rm tr}}

\begin{document}
	\title{Out of Time Order Correlations in the Quasi-Periodic Aubry-Andr\'e model}
	\author{Jonathon Riddell}
	\affiliation{Department of Physics \& Astronomy, McMaster University
		1280 Main St.  W., Hamilton ON L8S 4M1, Canada.}
        \author{Erik S. S{\o}rensen }
	\affiliation{Department of Physics \& Astronomy, McMaster University
		1280 Main St.  W., Hamilton ON L8S 4M1, Canada.}
	
	\date{\today}

\begin{abstract}
We study out of time ordered correlators (OTOC) in a free fermionic model with
a quasi-periodic potential. This model is equivalent to the Aubry-Andr\'e
model and features a phase transition from an extended phase to a localized phase at a
non-zero value of the strength of the quasi-periodic potential. 
We investigate five different time-regimes of interest
for out of time ordered correlators; early, wavefront, $x=v_Bt$, late time
equilibration and infinite time. For the early time regime we observe a power law for
all potential strengths. For the time regime preceding the wavefront 
we confirm a recently proposed universal form and use it to extract the characteristic velocity
of the wavefront for the present model. A Gaussian waveform is observed to work well in the time regime surrounding $x=v_Bt$. Our main result is for the late time equilibration regime where 
we derive a finite time equilibration bound for the OTOC, bounding the correlator's distance from its late time value.
The bound impose strict limits on equilibration of the OTOC in the extended regime
and is valid not only for the Aubry-Andr\'e model but
for any quadratic model. Finally, momentum out of time
ordered correlators for the Aubry-Andr\'e
model are studied where large values of the OTOC are observed at late times at the
critical point.
	
\end{abstract}

\maketitle

\section{Introduction}
	
Recently out of time ordered correlators (OTOCs) have experienced a resurgence of 
interest from different fields of physics ranging from 
the black hole information problem~\cite{Maldacena2016a} to information propagation in condensed matter
systems~\cite{yoshida2019,Swingle2017,Alonso2019,Yan2019,Tuziemski2019,Mao2019,Lewis-Swan2019,Nakamura2019}. 
The OTOC is of particular interest due to its role in witnessing the spreading
or``scrambling" of locally stored quantum information across all degrees of
freedom of the system, something traditional dynamical correlation functions of
the form $\langle A(t)B\rangle$ cannot. Thus, thermalization must have
information scrambling as a precursor since the thermal state necessarily will
have lost information about any initial state, although thermalization
typically occurs at a significantly longer time-scale~\cite{Bohrdt2017}.  An
upper bound for the initial exponential growth, $e^{\lambda_L t}$, of the OTOC,
with $\lambda_L\leq 2\pi k_B T/\hbar$ has been
conjectured~\cite{Maldacena2016a}. Models approaching or saturating this bound
are known as fast scramblers, in contrast to many condensed matter systems
which exhibit a much slower growth and are therefore known as slow scramblers.
The introduction of disorder significantly alters the information spreading,
restricting it within a localization length in Anderson
insulators~\cite{Riddell2} and partially halting the growth of the OTOC in
many-body localized states~\cite{Swingle2017}.  
The OTOC is directly related to the Loschmidt Echo~\cite{Yan2019}
and is has been established that the second Renyi entropy can be
expressed in terms of a sum over appropriately defined OTOCs~\cite{Fan2017}.
Any bound that can be established on the growth of the OTOC therefore implies a
related bound on the entanglement.  
A further understanding of the dynamics of
quantum information in models with both extended and localized states is
therefore of considerable interest and our focus here is on understanding how
this arises in the quasi-periodic Aubry-Andr\'e (AA) model where a critical
potential strength separates an extended and localized phase.

An OTOC is generally written in the form, 
\begin{equation}
\label{eq:defotoc}
C(x,t) = \langle [\hat{A}(t),\hat{B} ]^\dagger  [\hat{A}(t), \hat{B} ] \rangle,
\end{equation}
where $\hat{A}$, $\hat{B}$ are local observables which commute at $t=0$. If the
observables are both hermitian and unitary the OTOC can be re-expressed as, 
\begin{equation}
        \label{eq:unitotoc}
C(x,t) = 2-2 \Re [F(x,t)],  
\end{equation}
where,
\begin{equation}
F(x,t) = \langle \hat{A}(t)\hat{B}\hat{A}(t)\hat{B}\rangle.
        \label{eq:F}
\end{equation}
Often one refers to both $F$ and $C$ as the OTOC.
From a condensed matter perspective the OTOC is a measure of an operator
spreading its influence over a lattice, and quantifies the degree of
non-commutativity between two operators at different times. If the initially
zero $C(x,t)$ remains non-zero for an extended period of time we say the system
has scrambled.  A closely analogous diagnostic tool, capable of detecting
information scrambling, can be defined in terms of the mutual information
between two distant intervals~\cite{Alba2019}.

From a measurement perspective $F(x,t)$ can be understood as a series of
measurements. First acting on the state with operator $\hat{B}$ at $t=0$ and
evolving in time to $t>0$, then acting on the state with operator $\hat{A}$,
then evolving for time $-t<0$. The OTOC is then obtained by calculating the
overlap between the resulting state and the state that is first evolved by $t$,
then acted upon by $A$, then evolved by $-t$ and finally acted upon by $B$.
Typically in the context of the OTOC one uses $\langle \dots \rangle$
as the thermal average, often at infinite temperature, but studies in a
non-equilibrium setting starting from product states have also been done
\cite{Lee2018,Riddell2,Chen2017}. Out of time correlators have also sparked
experimental interest and significant progress has been made to reliably
measure these quantities
\cite{Swingle2016,Zhu2016,Yao2016,Danshita2017,Garttner2017}. The correlators
have even been reliably simulated on a small quantum computer \cite{Li2017} and
recently on an ion trap quantum computer~\cite{Landsman2019}.

The dynamics of the OTOC has five important regimes; early time, the wavefront, $x=v_Bt$,
late time dynamics and the infinite time limit.  The early time growth of OTOCs
has been of interest as an initial growth of the OTOC that precedes classical
information. If the Hamiltonian is local in interactions then use of the
Hadamard formula (see ref. \cite{wmillersymmetry} lemma 5.3) allows one to
conclude that in the early time regime the OTOC grows with a power law in time, 
\begin{equation}
	C(x,t) \sim t^{ l(x)}, 
\end{equation}
where $t$ is small and $l(x)$ is a linearly increasing function of the
distance. The early power law growth in time occurs before the wavefront hits
and is known to be independent of the integrability of the model
\cite{Dora2017,Roberts2016, Chen2018,LinOTOCising,Riddell2,Lee2018,Bao2019}.
This polynomial form is also known to be independent of disorder strength and
has been observed to hold in localized regimes \cite{Riddell2, Lee2018}. 

More interestingly, the wavefront tracks the passage of classical information in
the system. A universal wavefront form has been proposed
\cite{Xu2018b,Khemani20182}, valid for $t\ll x/v_B$,
\begin{equation} 
        \label{eq:universalform}
C(x,t) \sim \exp\left(-\lambda_L \frac{(x-v_Bt)^{1+p}}{t^p} \right),
\end{equation}
where $\lambda_L$ is the Lyapunov exponent and $v_B$ is the butterfly velocity.
Several other forms have been proposed, for a review see Ref.~\cite{Xu2018b}.
The above wavefront form, Eq.~(\ref{eq:universalform}), has been confirmed in several cases, and even used to show a
chaotic to many body localization transition
\cite{Nahum2018,Keyserlingk2018,Jian2018,Gu2017,Xu2018a,Xu2018b,Khemani20182,Sahu2018,Rakovsky2018,Shenker2014a,Patel2017,Chowdhury2017,Jian2018}.
For free models one can show with a saddle point approximation that the form
from Eq.~(\ref{eq:universalform}) takes $p = \frac{1}{2}$ and $v_B$ is the
maximal group velocity of the model \cite{Xu2018a,Xu2018b,Khemani20182}.
A particular appealing feature of Eq.~(\ref{eq:universalform}) is the appearance of a well-defined Butterfly velocity, $v_B$ for
a large range of models.
A recent numerical study focusing on the
random field XX-model suggested that for this disordered model a different form
could be made to fit better over an extended region~\cite{Riddell2} surrounding $x=v_B t$. This
result suggests further studies are important for understanding how quantum information is
spreading through the system.

The late time dynamics of OTOCs are a similarly rich regime of interest. 
Understanding how the function $g(t) = |C(x,t) - C(x,t\to \infty)|^2$ decays 
in time has received attention in many models. In the case of the anisotropic XY model the decay of the 
OTOC to its equilibrium value is an inverse power law \cite{LinOTOCising,Bao2019},
\begin{equation}
C(x,t) \sim \frac{1}{t^\alpha} +\gamma,
\end{equation}
where $\alpha \geq 0$ depending on the choices of spin operators and the
anisotropy, and $\gamma$ is the equilibrium value.  Other work has
been done on interacting systems where both inverse power laws were observed
for chaotic and many body localized phases, and even an exponential decay in
time for  Floquet systems~\cite{Chen2017,Swingle2017}. However, these results
are mostly numerical, and do not give rigorous bounds or arguments as to whether
or not the OTOC reaches equilibrium and if it does, to what resolution. 
Another aspect of the 
late time regime, the quantity $C(x,t\to \infty)$ in it self, is naturally of considerable
interest. In this setting $F(x, t \to \infty)$ is often chosen as the
quantity to study. In the presence of chaos we expect $F$ to
equilibrate to zero, and in other cases settle at a finite value between zero
and one
\cite{Huang2017,Fan2017,Chen2016,Swingle2017,Helu2017,Riddell2,LinOTOCising,Bao2019,Chen2017,Lee2018,Roberts2017,Huang2017v2,Chen2016v2,McGinley2018}.
A particularly important case for our purposes are the non-interacting models where
the observables defining the OTOC are both local in fermionic and spin representations on the
lattice. Here $F(x,t)$ is expected to initially decay towards zero, but
eventually return to $F(x,t) = 1$ and in the presence of disorder need not
decay back to its initial value or even equilibrate
\cite{Bao2019,Riddell2,LinOTOCising,Chen2016,McGinley2018}. Of course, $C$ is then 
predicted to follow the opposite behaviour, starting at zero then reaching a maximum.
It is also noteworthy that, in the proximity of a quantum critical point the OTOC has been shown to follow dynamical scaling laws~\cite{Wei2019}.

The introduction of disorder, with the potential of leading to localization, 
significantly changes the behaviour of the OTOC and propagation of quantum information
as a whole. Naturally, quantum information dynamics is expected to be dramatically different between
localized and extended phases. We therefore focus on the one-dimensional quasiperiodic 
Aubry-Andr\'e (AA) model~\cite{AubryAndre,Hiramoto1989}:
\begin{equation}
        H = -\frac{J}{2}\sum_j (|j\rangle\langle j+1|+\mathrm{h.c})+\lambda\sum_j\cos(2\pi\sigma j)|j\rangle\langle j|.
        \label{eq:AA1}
\end{equation}
Here, $J$ is the hopping strength and $\lambda$ the strength of the quasi-periodic potential.
This model has been extensively
studied~\cite{Aulbach2004,Boers2007,Modugno2009,Albert2010,Ribeiro2013,Danieli2015,Wang2017,Li2017b,Martinez2018,Castro2019} and since
it is quadratic large-scale exact numerical results can be obtained from the exact solution. In particular
quench dynamics has recently been studied~\cite{Gramsch2012}.
Crucially, it is well established that a critical potential strength
$\lambda_c=J$ separates an extended and localized regime if $\sigma$ is chosen
to be the golden mean $\sigma=(\sqrt{5}-1)/2$. For finite lattices this strictly
only holds if the system size is chosen as $L=F_i$, with $F_i$ a Fibonacci
number, and $\sigma=F_{i-1}/F_i$ approaching the golden mean as $i\to\infty$. A
dual model can then be formulated~\cite{AubryAndre,Aulbach2004} by
introducing the dual basis $|\bar k\rangle = L^{-1/2}\sum_j\exp(i2\pi\bar k\sigma j)|j\rangle$.  
$\lambda_c=J$ is then the self-dual point. The extended phase is characterized by ballistic transport
as opposed to diffusive~\cite{AubryAndre}. The nature of the quasi-periodic potential is also special since
no rare regions exists and it has recently been argued that localization in the AA model is fundamentally more classical
than disorder-induced Anderson localization~\cite{Albert2010}.
It is possible to realize this model quite closely 
in optical lattices and studies of both bosonic and fermionic experimental realizations have been pursued using 
$^{39}$K bosons~\cite{Roati2008,Deissler2010,Lucioni2011},
$^{87}$Rb bosons~\cite{Fallani2007}, and $^{40}$K fermions~\cite{Schreiber2015,Luschen2017a,Luschen2017b}.

The AA model has also recently been studied in the presence of an interaction
term~\cite{Iyer2013,Xu2019a}.  While no longer exactly solvable, a many-body
localized phase can be identified in studies of small
chains~\cite{Iyer2013,Xu2019a} and by analyzing the OTOC it has been suggested
that an intermediate 'S' phase occurs between the extended and many-body
localized phases with a power-law like causal lightcone~\cite{Xu2019a}.

The structure of this paper is as follows, in section \ref{sec:modelintro} we
discuss our formulation of the Aubry-Andr\'e model and describe the quench
protocol we use. In section \ref{Sec:posOTOC} we investigate the dynamics of an
out of time ordered correlator in real space and break the section into three
subsections dedicated to three dynamical regions of interest. In subsection
\ref{sec:early} we show that when quenching into either the extended,
localized, or critical phase a power-law growth is observed in the early time
regime. In \ref{sec:wavefront} we investigate the discrepancies between
\cite{Xu2018b,Khemani20182} and \cite{Riddell2} for times closer to the
wave-front. Section \ref{sec:latetime} contains a proof that, in the extended
phase of a free model, we expect the out of time ordered correlator to
equilibrate even in the presence of the quasi-periodic potential.  The infinite
time value is also shown to be {\it zero} regardless of the strength of the quasi-periodic potential indicating a
lack of scrambling regardless of disorder in the extended phase. Finally in
section \ref{sec:momentum} we investigate OTOCs constructed from momentum
occupation operators and find that they obey a simple waveform.

\section{The model and OTOCs} \label{sec:modelintro}
As outlined, we focus on the quasi-periodic AA model. We chose a fermionic representation and write the Hamiltonian as follows:
\begin{equation}
		\hat{H} = \sum_{i,j}M_{i,j}\hat{f}_i^\dagger \hat{f}_j,
\end{equation}
where the effective elements of the Hamiltonian matrix $M$ is filled by, $M_{i,j} =
-\frac{J}{2}$ if $|i-j| =1$ and $M_{j,j} = \lambda \cos(2\pi \sigma j)$.  The
operators are fermionic so we have  $\{\hat{f}_k,\hat{f}_l \}
=\{\hat{f}_k^{\dagger},\hat{f}_l^{\dagger}\}= 0$ and
$\{\hat{f}_l^{\dagger},\hat{f}_k\} = \delta_{l,k}$. 
All other entries of the effective Hamiltonian are zero.
Note that this corresponds to open boundary conditions with nearest neighbour hopping which is the most
convenient for the calculations.
The constant $\sigma$ is the inverse golden ratio, $\sigma = (\sqrt{5}-1)/2$. 
For the very large system sizes we use we have not been able to observe any numerical difference
between using $L=F_i,\ \sigma=F_{i-1}/L$ and using a large $L$ with $\sigma=(\sqrt{5}-1)/2$ even though
the model is strictly no longer self-dual. For convenience we therefore use the latter approach.
Since the inverse golden ratio is irrational, this creates a quasi-periodic potential controlled by the
value of $\lambda$.  For the rest of our discussion we set $J = 1$ and $\hbar = 1$. 
This model is identical to the Aubry-Andr\'e model as can be seen through a series of
transformations \cite{AubryAndre,Coleman}. One can easily diagonalize
and time evolve states in this model, the details of which are presented in the
Appendix~\ref{app:tevolve}. As described above, this model is known to have a localization transition at a critical
point $\lambda_c = J$. For $\lambda < \lambda_c$ all states are extended, and
$\lambda > \lambda_c$ all states are localized with localization length $\xi =
\frac{1}{\ln \lambda}$ \cite{AubryAndre}. Relaxation and thermalization following a quench
into both extended and localized phases has recently been 
investigated in this model~\cite{Gramsch2012}. While most one-body observables thermalize to a generalized Gibbs ensemble
in the extended state, and some in the localized, special dynamics was observed for a quench to the critical points where
the observables investigated did not reach a clear stationary value in the time
intervals investigated.  Similar quadratic fermionic models have been used to
investigate OTOCs at large system sizes, showing non-trivial behaviour of both
non-disordered and disordered OTOC investigations  in integrable
models \cite{Riddell2,LinOTOCising,Muralid}. 

The OTOCs we will be interested in are written in the form Eq.~(\ref{eq:defotoc})
where we choose $\hat{A}$ and $\hat{B}$ such that they commute at $t=0$ and are unitary. 
The operators being hermitian and unitary then obey Eq.~(\ref{eq:unitotoc}), (\ref{eq:F}).
In general we choose our operators such that at $t=0$ $[\hat{A}, \hat{B}] = 0$,
making $C(x,0) = 0$ in all cases. This gives us a convenient reference point in
time. Because we are talking about fermionic operators, it makes sense to only
consider operators which are quadratic, and further, we choose to restrict
ourselves to operators that can be expressed as number operators in real or momentum space.
In momentum space the operators we consider are :
\begin{eqnarray}
\eta_k := \frac{1}{\sqrt{L}} \sum_j e^{ikj} \hat{f}_j \\
\eta_k^\dagger := \frac{1}{\sqrt{L}} \sum_j e^{-ikj} \hat{f}_j^\dagger.
\end{eqnarray}
Where, $k\in 2\pi m/L$ with $m = 1,2 \dots L$. 
These operators are extremely non-local in the real space operators, and for
the case of $\lambda=0$ and periodic boundary conditions, are the operators
which diagonalize $M$ (strictly speaking only when periodic boundary conditions are used). 
It has been observed previously that operators not local
in the fermionic representation show fundamentally different behaviour than
the local ones~\cite{LinOTOCising}. These however were spin operators, which
were non-local in the Jordan-Wigner transform, so investigating OTOCs with
momentum number operators is not entirely an exact analogue. 

\section{Real space OTOCs} \label{sec:realspace}
	\label{Sec:posOTOC}
We start by considering OTOCs based on operators defined in real space.
To be specific we study the following operators, 
\begin{equation}
\hat{A}(t) = 2\hat{f}_{\frac{L}{2}}^\dagger(t)\hat{f}_{\frac{L}{2}}(t)-1 \enspace \text{,} \enspace \hat{B} = 2\hat{f}_{j}^\dagger\hat{f}_{j}-1
\end{equation}
Where we have fixed the location of $\hat{A}$ in space at the middle point of
the lattice, and we will vary the location of $\hat{B}$, so we see can observe
the effect of $\hat{A}$ spreading over the lattice. The operators are written
with a factor of $2$ and a subtraction of $1$ to make them unitary. The
dynamics and calculations of the OTOC in this setting is presented in 
Appendix~\ref{app:tevolve}, \ref{app:quenchprotocol} and \ref{app:calcotoc}.
	
\subsection{Early time} \label{sec:early}
In this section we explore the early time behavior of the real space OTOCs. As seen in Eq. \ref{eq:OTOCeq} the dynamics of the OTOC are dominated by the squared anti-commutator relation of the fermionic operators in time, $a_{m,n}(t)$ (defined in Eq. \ref{eq:A8}). If one sets $\lambda = 0$ and assumes periodic boundary conditions, one finds that
in the thermodynamic limit that the squared anti-commuter behaves as the square of a Bessel
function in time (see for example Appendix C of\cite{Xu2018b}), 
\begin{equation}
C(x,t) \sim |a_{m,n}(t)|^2 \sim J_x^2(t),
        \label{eq:Bessel}
\end{equation}
then in the limit of small $t$ one finds that, 
\begin{equation} \label{eq:early}
C(x,t) \sim t^{2|x|}.
\end{equation}

\onecolumngrid

\begin{figure}[t]
\centering
\includegraphics[width=\linewidth]{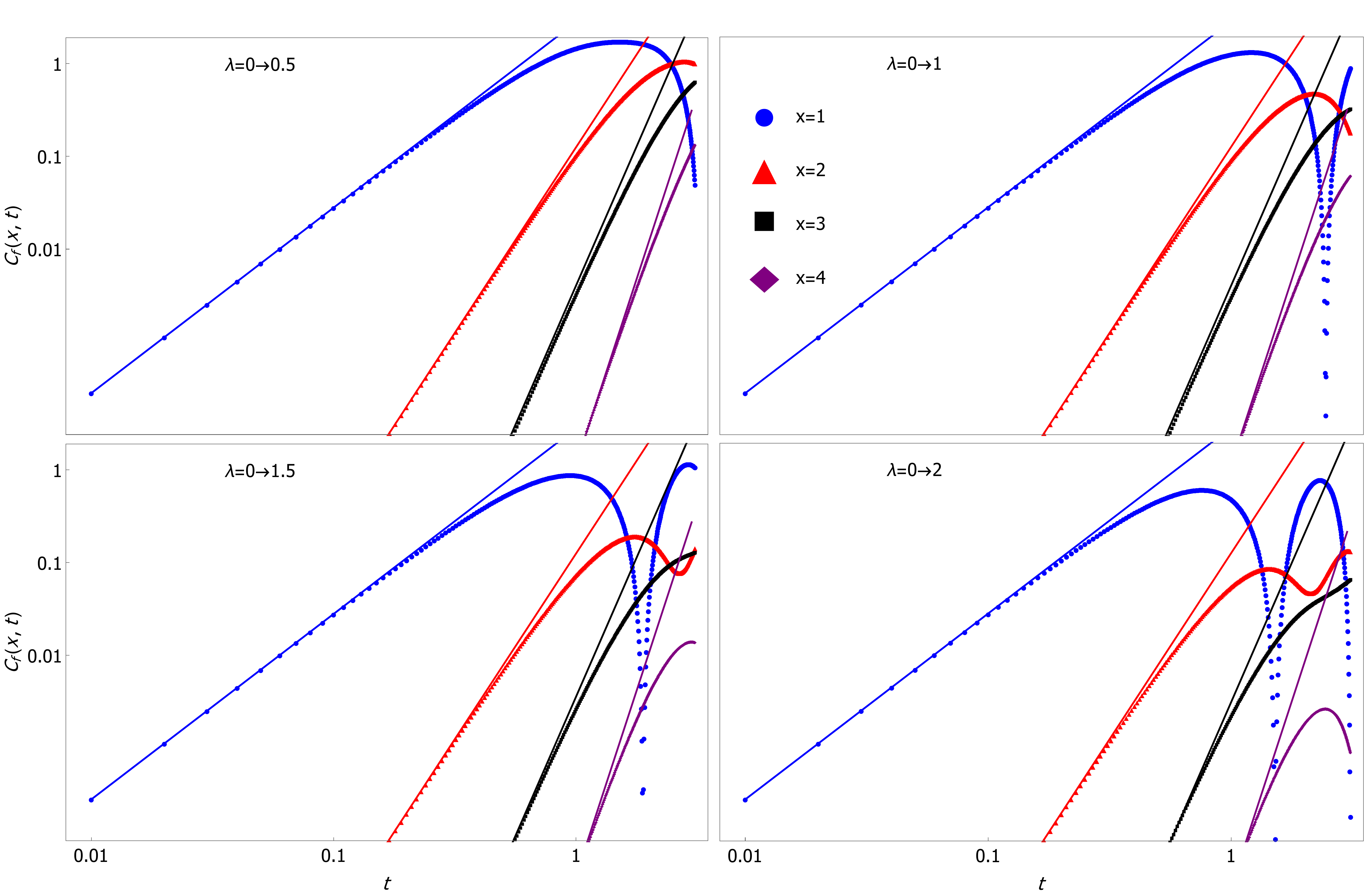}
\caption{Early time behavior of $C(x,t)$ at different distances. 
        The solid curve is the power law and the dotted curved is the data collected for the OTOC. 
        The system size is $L=1200$. Results are shown for quenches to four different values of $\lambda$ starting
        from the ground state of the model at $\lambda = 0 $.}
\label{fig:early1}
\end{figure}

\twocolumngrid

For our purposes the derivation sketched above is too restrictive as we are
also interested in non-translationally invariant models and our OTOC features more dynamical terms than just the squared anti-commuter. However, the result, Eq.~(\ref{eq:early}) still remains
correct even in the presence of non-zero quasi-periodic potential. This can be seen through the use of
the Hadamard formula as shown in \cite{Riddell2}.

We study this prediction in the most dynamically rich way possible, by
quenching from the half-filled ground-state at $\lambda=0$ to $\lambda = 0.5,1,1.5,2$. Our
results are shown in Fig.~ \ref{fig:early1}.  For a detailed discussion of the starting state see
Appendix~\ref{app:quenchprotocol}.
The results here do not significantly change if the quench
is to the localized phase ($\lambda =1.5,2$), critical ($\lambda=1)$ or extended phase ($\lambda=0.5$). For
all strengths of the quasi-periodic potential is a power-law behaviour observed following Eq.~(\ref{eq:early}). This results agree with \cite{Riddell2} which found that in an Anderson localized model regardless of the strength of the 
localization, if the OTOC significantly grows, then the polynomial early time growth Eq.~(\ref{eq:early})
is observed to be hold. 
This follows naturally from the fact that Eq.~(\ref{eq:early})
is independent of the potential strength, the first contributing dynamics to the OTOC are unaffected by the potential term
and come solely from the hopping
terms. The early time behaviour can therefore be obtained by studying the $\lambda=0$ case. 

\subsection{Wavefront} \label{sec:wavefront}
In this section we study the wavefront at different potential strengths and address
discrepancies from the results shown in  \cite{Xu2018b,Khemani20182} and
\cite{Riddell2}. Recently, the universal form was claimed to be confirmed in
the XX spin chain, contradictory to earlier claims \cite{Bao2019}. Here we
discuss these seemingly contradictory claims. The universal wave form
predicted for the out of time ordered correlator in free theories by means of a
standard saddle point approximation scheme is given by Eq.~(\ref{eq:universalform}) in terms of
the Lyapunov exponent, $\lambda_L$ and the Butterfly velocity, $v_B$.
Often this form is applied at surprisingly early times~\cite{Xu2019a} where $-50<\log(C)<-10$.
For the AA model with $\lambda = 0$, corresponding to free fermions, we expect the $v_B = J$ as the maximal group
velocity, and $p = \frac{1}{2}$. The universal form,
Eq.~(\ref{eq:universalform}), cannot be re-expressed in a form equivalent to
the 'Gaussian' form characterized by two spatial and disorder dependent functions $a(x,\lambda)$, $b(x,\lambda)$ proposed in Ref.~\onlinecite{Riddell2}, for times surrounding $x=v_Bt$, for a fixed $x=x_0$:
\begin{equation}
\label{eq:ourfit} C(x=x_0,t) \sim e^{-a(x,\lambda)\left(\frac{t^2}{2}-\frac{xt}{v_B}\right)+b(x,\lambda)t}.
\end{equation} 
We can rewrite Eq. \ref{eq:ourfit} as, 

\begin{equation} \label{eq:Gaussian}
	C(x=x_0,t) \sim e^{ -m(x,\lambda)\left(t-\frac{x}{v_B} \right)^2+b(x,\lambda)t},
\end{equation}
where $m(x,\lambda) = a(x,\lambda)/2$.

We expect that the discrepancy is most likely due the existence two unique time regimes that are close together. To eliminate noise in our OTOC we drop all of the dynamical terms except the squared anti-commutator. which is  equivalent to instead studying the OTOC, 
\begin{equation}
        \label{eq:nonoiseOTOC}
	C(x,t) = \tr \left( \{\hat{f}_m^\dagger(t),\hat{f}_n \} \{\hat{f}_m(t),\hat{f}_n^\dagger  \}   \right) \equiv |a_{m,n}(t)|^2.
\end{equation}

\begin{figure}[h!]
\centering
	\includegraphics[width=\linewidth]{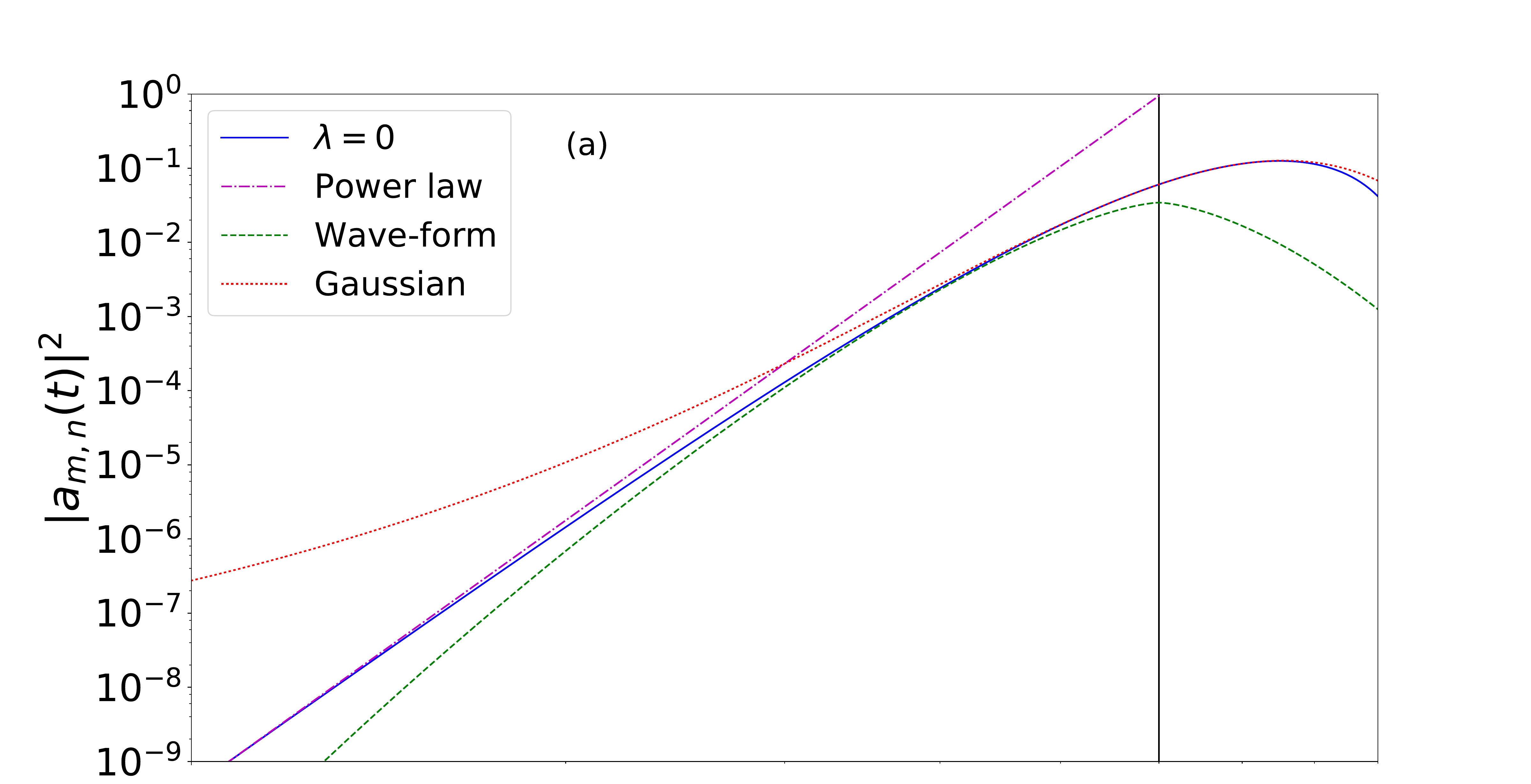}
	\includegraphics[width=\linewidth]{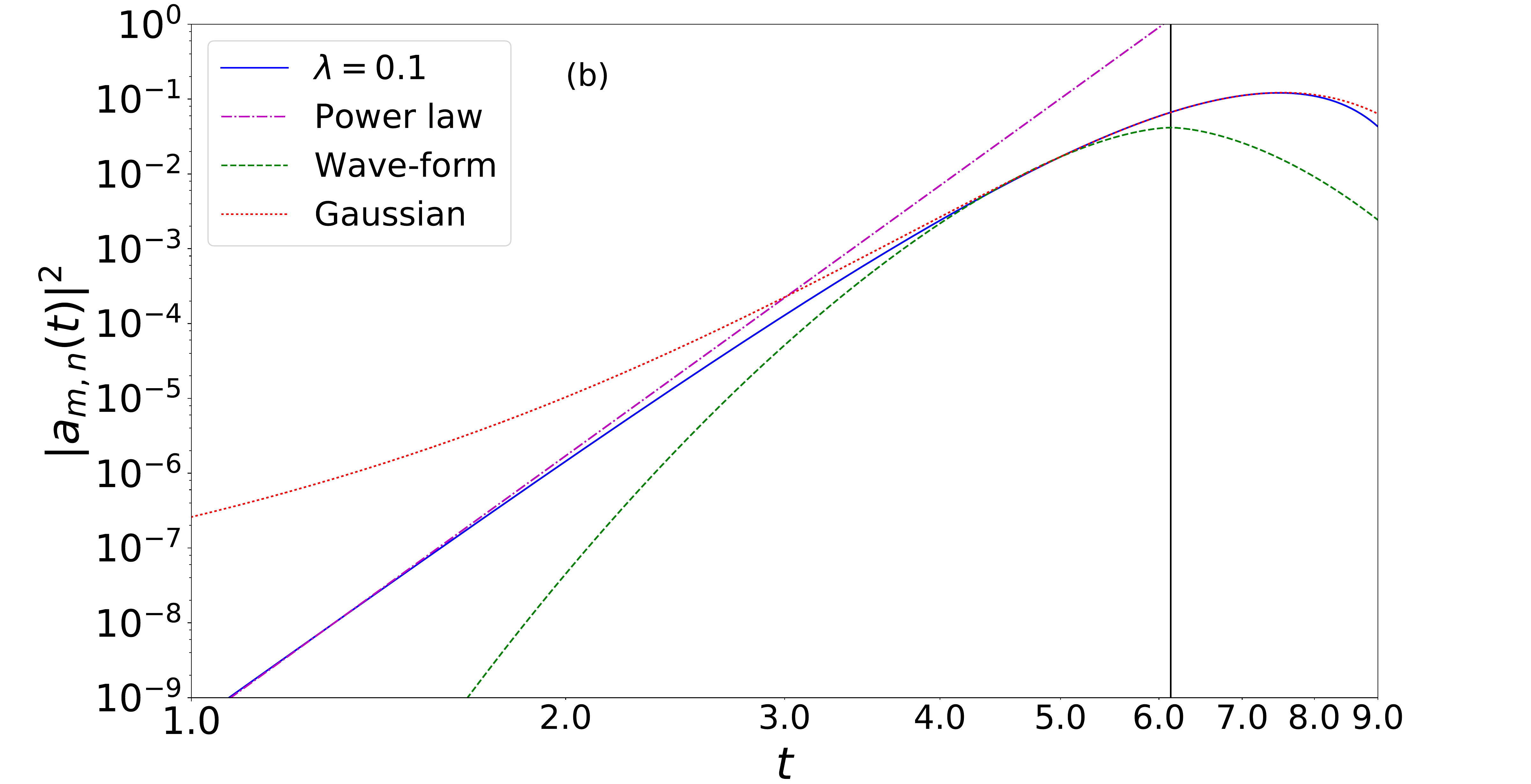}
        \caption{$|a_{m,n}(t)|^2$ for $\lambda=0$ (a) and $\lambda=0.1$ (b) plotted with the fitting functions of the early time form, Eq.~(\ref{eq:early}),
        the proposed universal wave form, Eq.~(\ref{eq:universalform}) and the Gaussian form, Eq.~(\ref{eq:Gaussian}).
        Results are for a fixed $x=6$ with $L=1600$ and $\lambda=0$. The vertical solid line in both panels corresponds the arrival
        of the classical wave front at $t=x/v_B$ using the fitted $v_B$. 
        }
	\label{fig:wavefront01}
\end{figure}
To further facilitate the analysis we include a phase, $\phi$, in the potential
$\lambda\cos(2\pi\sigma j+\phi)$ and smooth our data by averaging over $\phi$.
Our results are shown in Fig.~\ref{fig:wavefront01} where we follow an analysis
similar to \cite{Sahu2018}.  By varying both time and space we fit the OTOC for
$\lambda = 0$ in the region such that $\log \left(|a_{m,n}|^2\right)\in[-10,-6]
$. With this fit we find $v_B=0.9950 \pm 0.0002$, $p=0.50\pm 0.08$ and
$\lambda_L=1.78 \pm 0.03$ for the universal form,
Eq.~(\ref{eq:universalform}). Where the errors reported are one standard
deviation of the parameter estimate. These values are in close agreement with
the expected values of $v_B = 1$ and $p = \frac{1}{2}$. Similarly we
investigated the $\lambda = 0.1$ case for $\log
\left(|a_{m,n}|^2\right)\in[-12,-8] $ and found $v_B=0.9783 \pm 0.0003$,
$p=0.647 \pm 0.03$ and $\lambda_L=2.153 \pm 0.09$ for the universal form,
Eq.~(\ref{eq:universalform}). However, these fits correspond to times that
significantly precede the classical wavefront. 
For larger values of the potential strength, $\lambda$, we have found it more
difficult to obtain good fits to the universal form,
Eq.~(\ref{eq:universalform}).

At later times the OTOC enters
a dynamical regime where the Gaussian form of Eq. \ref{eq:Gaussian} is valid.
Fixing $x = 6$ and using the $v_B$ found for the universal form we find that for $\lambda = 0$ $m(x,\lambda) = 0.3027 \pm
0.0001$  and $b(x,\lambda) = 0.9470 \pm 0.0001$. For $\lambda = 0.1$ we find
$m(x,\lambda) = 0.3052 \pm 0.0001$ and $b(x,\lambda) = 0.8597 \pm 0.0002$.

\begin{figure}[h!]
\centering
	\includegraphics[width=\linewidth,clip=true]{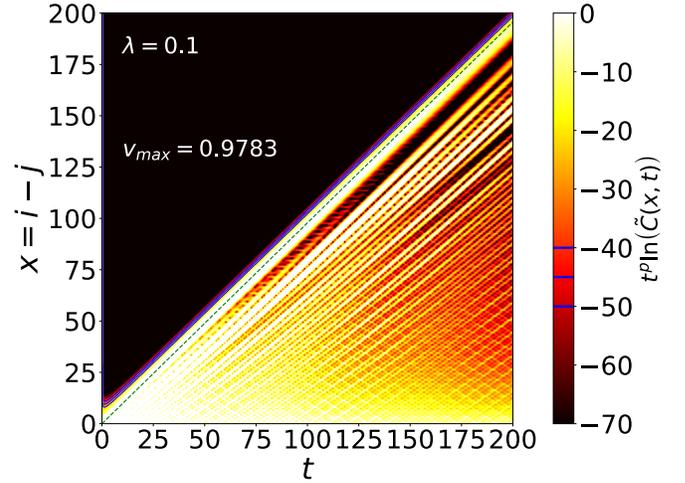}
        \caption{Density plot of $t^p\log(\tilde C)$, with $\tilde C$ an appropriately normalized OTOC from Eqn. \ref{eq:nonoiseOTOC} and $p=0.6470$. 
        Results are shown for $\lambda=0.1$ and $L=1600$. Contour lines are plotted as solid blue lines. The dashed green
        line indicates $x=v_B t$ with $v_b=0.9783$ which appear closely parallel to the contour lines.
        }
	\label{fig:wavefrontp1}
\end{figure}
To further illustrate the universal form, Eq.~(\ref{eq:universalform}), we show in Fig.~\ref{fig:wavefrontp1} results for the entire
$C(x,t)$ over a large range of $x$ and $t$ for $\lambda=0.1$. As above we have smoothened the data over the phase $\phi$. We first appropriately normalize 
$C$ to obtain $\tilde C$ and then plot $t^p \log (\tilde C)$ using the fitted $p=0.6470$. 
We then expect that contour lines should be straight lines defined by $x=v_b t$. This is clearly observed in Fig.~\ref{fig:wavefrontp1} although
we note that it is only contour lines for extremely small values of $t^p \log (\tilde C)$ (of the order of $-40$ to $-50$) that are completely parallel
to the determined $v_B t$.  Although the universal form of Eq.~\ref{eq:universalform} seems to work well, it is only applicable at times $t\ll \frac{x}{v_B}$. 

Let us return to the Gaussian form of Eq. \ref{eq:Gaussian}, expected to be valid close to $x=v_Bt$. We consider the
behaviour of the functions $m(x,\lambda)$ and $b(x,\lambda)$ by varying $x$ and fixing $v_B$ as the velocities found fitting the universal waveform. These functions appear to asymptotically approach a fixed value in
the large $x$ limit. 
\begin{figure}[h!]
	\centering
	\includegraphics[width=\linewidth]{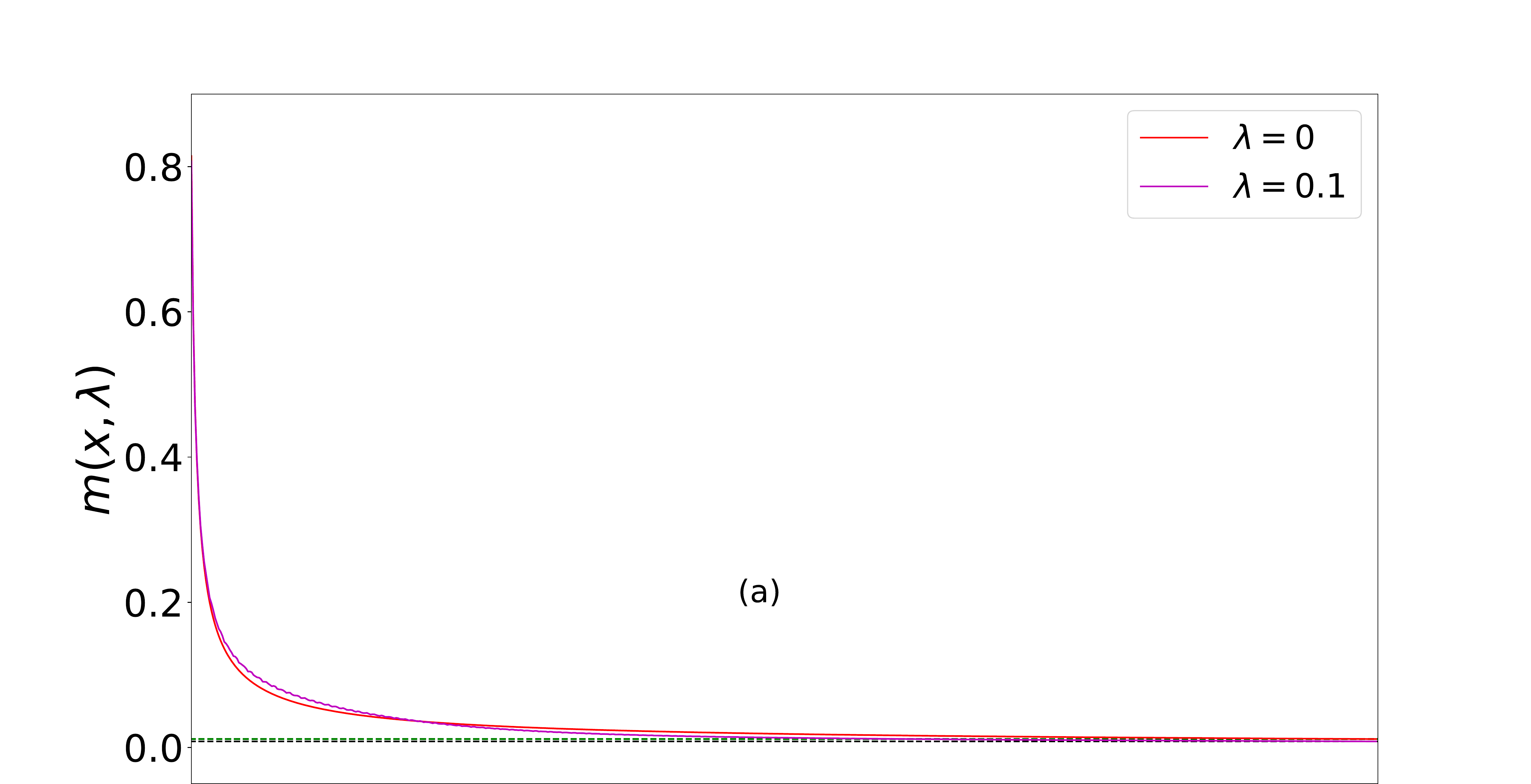}
	\includegraphics[width=\linewidth]{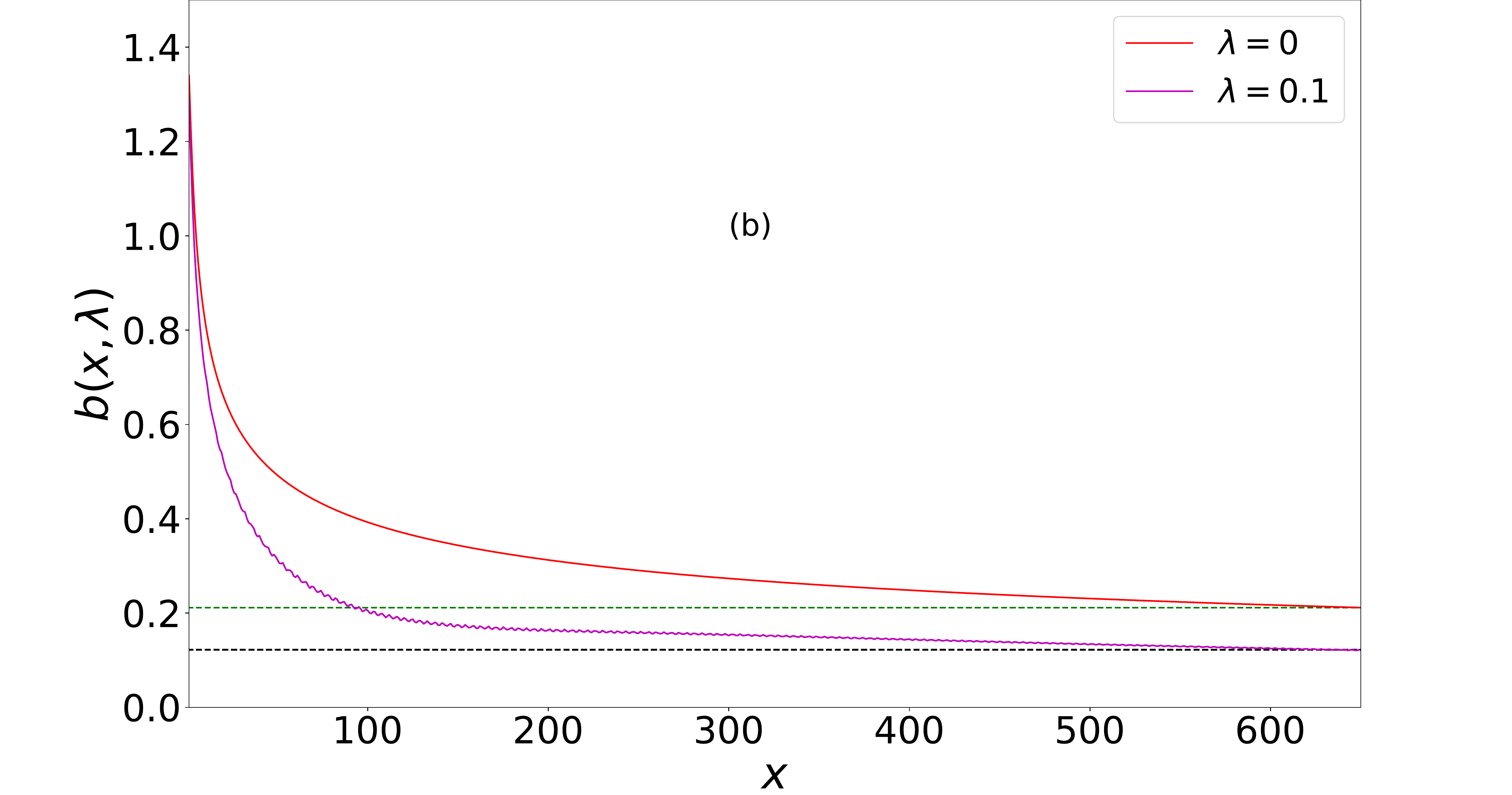}
	\caption{Functions $m(x,\lambda)$ (a) and $b(x,\lambda)$ (b) behaviour for fixed $\lambda$ at different $x$. Results are shown for no $\phi$ averaging and $L = 1600$. The dashed green horizontal line corresponds to the observed value of the function at $x=650$ for $\lambda = 0$ and the dashed black line to the value for $\lambda = 0.1$.
	}
	\label{fig:largex}
\end{figure}
For large $x$ and $\lambda = 0$  $m(x,\lambda) \approx 0.01$ and $b(x,\lambda) \approx 0.2$. For $\lambda = 0.1$ we see the values   $m(x,\lambda) \approx 0.008$ and $b(x,\lambda) \approx 0.1$. This result is shown in Fig. \ref{fig:largex}. Errors on this parameters are on the order of $10^{-4}$ or smaller. This means that taking large values of distance between the two observables $\hat{A}$ and $\hat{B}$, we may write, 

\begin{equation}
	C(x,t) \sim e^{-m(t-\frac{x}{v_B})^2}e^{bt},
\end{equation}
where $m$ and $b$ are positive constants. Intuitively this corresponds to a Gaussian wave travelling at velocity $v_B$, augmented by $e^{bt}$. This form is expected to be valid on the interval surrounding the passage of classical information around $x=v_Bt$. Hence, this form for works rather close to $x=v_Bt$. It seems likely that in interacting systems this might be apparent for much smaller values of $x$. 

\begin{figure}[h!]
	\centering
	\includegraphics[width=\linewidth]{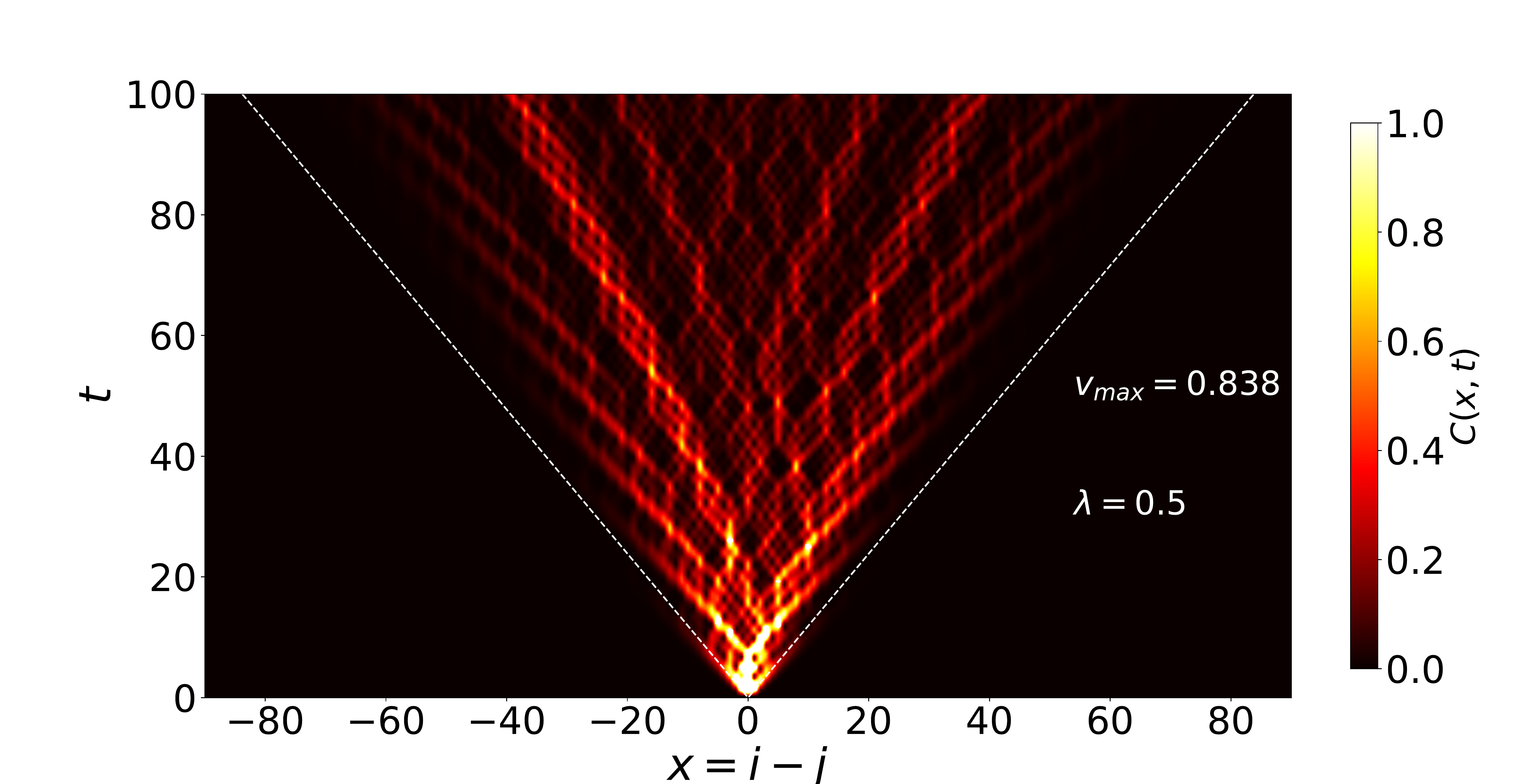}
	\caption{Wavefront spreading in both $x$ and $t$ for $\lambda = 0.5$, 
        the center being taken as $\frac{L}{2}$. System size $L=1200$.}
	\label{fig:thermwavefrontp5}
\end{figure}
If we instead of using the OTOC defined from the anti-commutator, Eq.~(\ref{eq:nonoiseOTOC}), use the full $C(x,t)$ with a thermal average where we fixed the inverse temperature $\beta = 1$
we find typical results as shown 
in Fig.~\ref{fig:thermwavefrontp5} for $\lambda=0.5$. In this case, as is this case for the remainder of our results
we do not smoothen the data using the phase $\phi$. From Fig.~\ref{fig:thermwavefrontp5} we see that the velocity 
predicted from the universal fit, Eq.~(\ref{eq:universalform}), of $v_B=0.838$
seems to be a good fit for predicting the spread of classical information.  For
larger values of $\lambda$ we have not found it possible to use the universal
form Eq.~(\ref{eq:nonoiseOTOC}) in contrast to recent results by Xu et
al~\cite{Xu2019a}. A possible explanation for this is that Xu et al
~\cite{Xu2019a} study the behaviour of the OTOC in a thermal state at infinite
temperature in an interacting model, a somewhat different setting.


\subsection{Late time} \label{sec:latetime}
It is also interesting to investigate the late time dynamics of the OTOC. In
prior studies it was pointed out that a $C(x,t) \sim \frac{1}{t}$ behaviour was
expected in late time \cite{Bao2019, LinOTOCising}. These results however are
for disorder-free models and do not in general hold for our discussion. So
instead we look to analytically show that these OTOCs indeed go to an
equilibrium value in the late time regime in the extended phase, regardless of strength of the quasi-periodic potential.
To bound this behaviour and prove equilibration we again focus on studying the OTOC defined in terms
of the squared anti-commutator, Eq.~(\ref{eq:nonoiseOTOC}). From Eq.~(\ref{eq:A8}) this can be written as:
\begin{eqnarray}
C(x,t) &=&\tr \left( \{\hat{f}_m^\dagger(t),\hat{f}_n \} \{\hat{f}_m(t),\hat{f}_n^\dagger  \}   \right) \equiv \left|a_{m,n}(t)\right|^2 \nonumber\\
        &=&\sum_{k,l}A_{m,k}A_{n,k}A_{m,l}A_{n,l} e^{i\left(\epsilon_{k}- \epsilon_{l}\right)t}. 
        \label{eq:at}
\end{eqnarray}
The infinite time average is defined as, 
\begin{equation}\label{eq:inftime}
\left|\omega_{m,n} \right|^2 = \lim_{T \to \infty} \frac{1}{T} \int_{0}^T |a_{m,n}|^2dt,
\end{equation}
using the fact that $\epsilon_{k} = \epsilon_{l} \Leftrightarrow k=l $,
\begin{equation} \label{eq:inftime2}
		\left|\omega_{m,n}\right|^2 = \sum_k A_{m,k}^2 A_{n,k}^2.
\end{equation} 
From Eq.~(\ref{eq:inftime2}) we can come to the intuitive conclusion that when the system is extended, in the thermodynamic limit $L\to\infty$, 
we expect the infinite time average to go to zero. The argument for this is as follows. In the extended phase, 
the values of $A_{m,k}$ will go like $A_{m,k} \sim \frac{1}{\sqrt{L}}$. Which leads to,
\begin{equation}
		\left|\omega_{m,n}\right|^2 \sim \frac{1}{L},
\end{equation}
approaching zero in the thermodynamic limit.
This is opposed to the localized phase where we expect, $A_{m,k} \sim e^{-|k-m|/\xi}$, with $\xi$ the localization length and $k = 1, \dots L$ \cite{RahmanXY} (see lemma 8.1). 
This makes the infinite time average go like, 
\begin{equation}
		|w_{m,n}|^2 \sim \max_k e^{-(|k-m|+|k-n|)/\xi}.
\end{equation}
Hence, the infinite time average of the OTOC is in this case non-zero within a distance of the order of the localization length.

Next we focus on bounding the relaxation process in time,
following \cite{Malabarba14,Garcia-PintosPRX2017,Riddell3}. To study the relaxation we define the positive function, 
\begin{equation} \label{eq:gfunc}
		g_{m,n}(t) = \left| \left|  a_{m,n}(t) \right|^2   - \left|\omega_{m,n}\right|^2 \right|^2.
\end{equation}
Eq.~(\ref{eq:gfunc}) can be interpreted as the distance the OTOC is from its
late time value, assuming such a value exists. To be precise we will work with
the time average of the function,
\begin{equation} \label{eq:timeavg}
		\langle g_{m,n}(t) \rangle_T = \frac{1}{T} \int_{0}^T \left|  \sum_{k\neq l}A_{m,k}A_{n,k}A_{m,l}A_{n,l} e^{i\left(\epsilon_{k}- \epsilon_{l}\right)t}  \right|^2 dt,
\end{equation}
to make notation easier let $\alpha = (k,l)$ and, 
\begin{equation} \label{eq:vdef}
		v_{\alpha} = A_{m,k}A_{n,k}A_{m,l}A_{n,l}, \enspace G_{\alpha} = \epsilon_{k}- \epsilon_{l}.
\end{equation}
This allows us to instead write the expression as, 
\begin{equation}
			\langle g_{m,n}(t) \rangle_T = \frac{1}{T} \int_{0}^T \sum_{\alpha , \beta}v_{\alpha} v_{\beta} e^{i\left(G_{\alpha}- G_{\beta} \right)t}dt,
\end{equation} 

We make use of the triangle inequality to make all elements of the sum positive, and then normalize, defining, $Q = \sum_\alpha |v_\alpha|$, 
\begin{equation}
		\langle g_{m,n}(t) \rangle_T \leq  Q^2 \frac{1}{T} \int_{0}^T \sum_{\alpha , \beta}p_{\alpha} p_{\beta}e^{i\left(G_{\alpha}- G_{\beta} \right)t}dt. 
\end{equation}
It is important to consider how big $Q$ might be. Trivially,
$Q \leq \sum_\alpha \max_\alpha |v_\alpha|$. 
Since this sum over $\alpha$ is quadratic in $L$ and restricting ourselves to the extended regime,  Eq.~(\ref{eq:vdef}) gives,  
$\max_\alpha |v_\alpha| \sim \frac{1}{L^2}$, it then follows that  $Q = O(1)$.

We now introduce the function, 
\begin{equation}
		\xi_p(x) = \max_{\beta} \sum_{\alpha:G_\beta \in [G_\beta, G_\beta+x]} p_\alpha.
\end{equation}
In Appendix $\ref{app:bound}$ we show that the time average can be bounded using a Gaussian profile, giving, 

\begin{equation} \label{eq:bound1}
\langle g_{m,n}(t) \rangle_T \leq  	\kappa \pi Q^2 \xi_p\left(\frac{1}{T}\right),
\end{equation}
where $\kappa \approx 2.8637$. To further bound this we introduce the two functions,

\begin{equation}
		a(\epsilon) = \frac{\xi_p(\epsilon)}{\epsilon}\sigma_G, \enspace \delta(\epsilon) = \xi_p(\epsilon),
\end{equation}
where $\sigma_G = \sqrt{\sum_\alpha p_\alpha G_\alpha^2 - \left(  p_\alpha
G_\alpha\right)^2}$ is the standard deviation of our distribution of
frequencies. From here on we assume $a(\epsilon)$ and $\delta(\epsilon)$ are implicitly dependent on $m,n$. It can be shown that (proposition 5 of ~\cite{Garcia-PintosPRX2017}):
\begin{equation}
		\xi_p(x) \leq \frac{a(\epsilon)}{\sigma_G}x+ \delta(\epsilon).
                \label{eq:xi}
\end{equation}
Using Eq.~(\ref{eq:xi}) we can rewrite Eq. \ref{eq:bound1} as:
\begin{equation} \label{eq:finitetime}
		\langle g_{m,n}(t) \rangle_T \leq \kappa \pi Q^2 \left( \frac{a(\epsilon)}{\sigma_GT}+ \delta(\epsilon)  \right).
\end{equation}
Eq. \ref{eq:finitetime} allows us to upper bound the time scale at which the OTOC equilibrates as $T_{eq} = \frac{ \kappa \pi a(\epsilon) Q^2}{\sigma_G}$.
\begin{figure}[t]
\centering
\includegraphics[width=\linewidth]{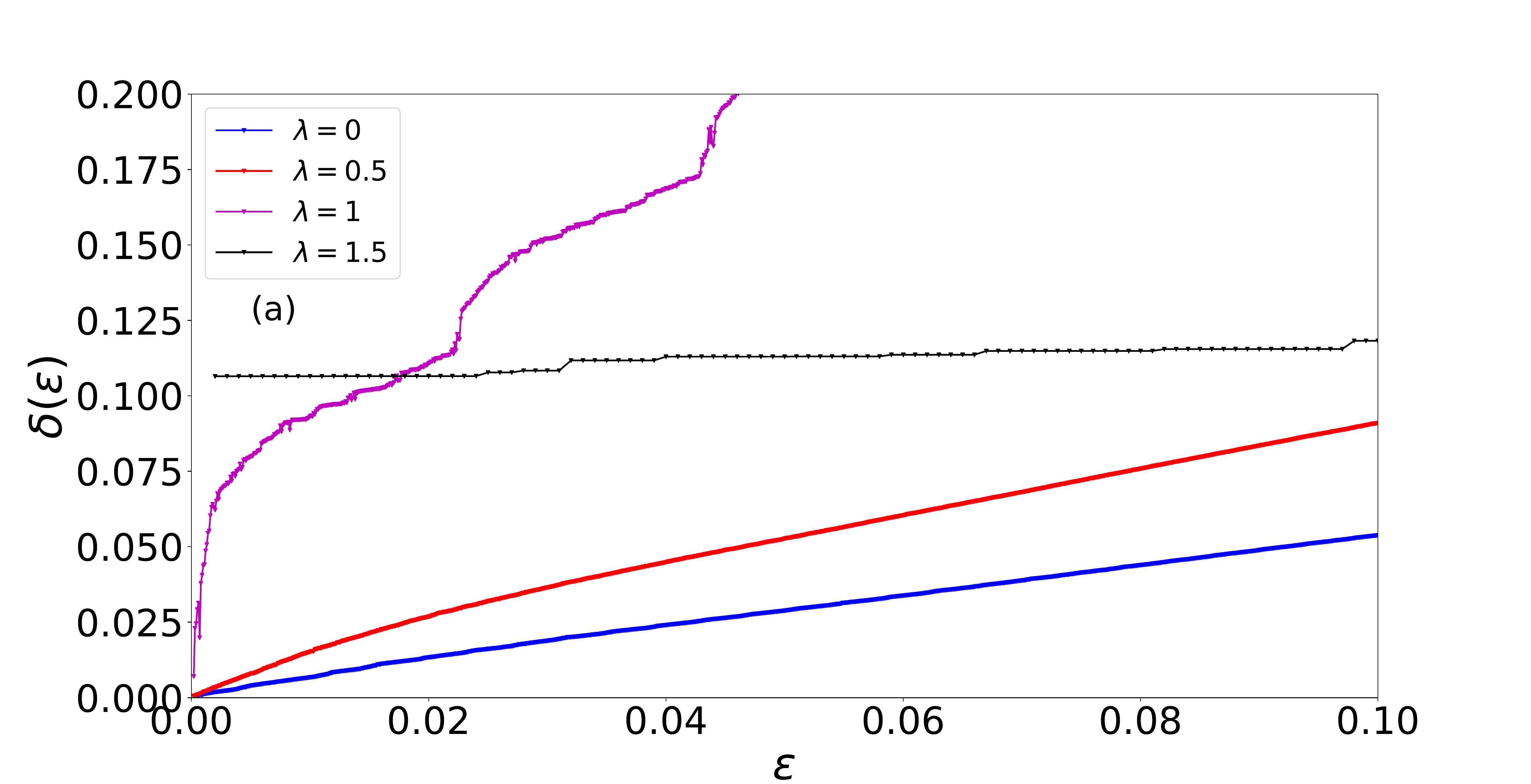}
\includegraphics[width=\linewidth]{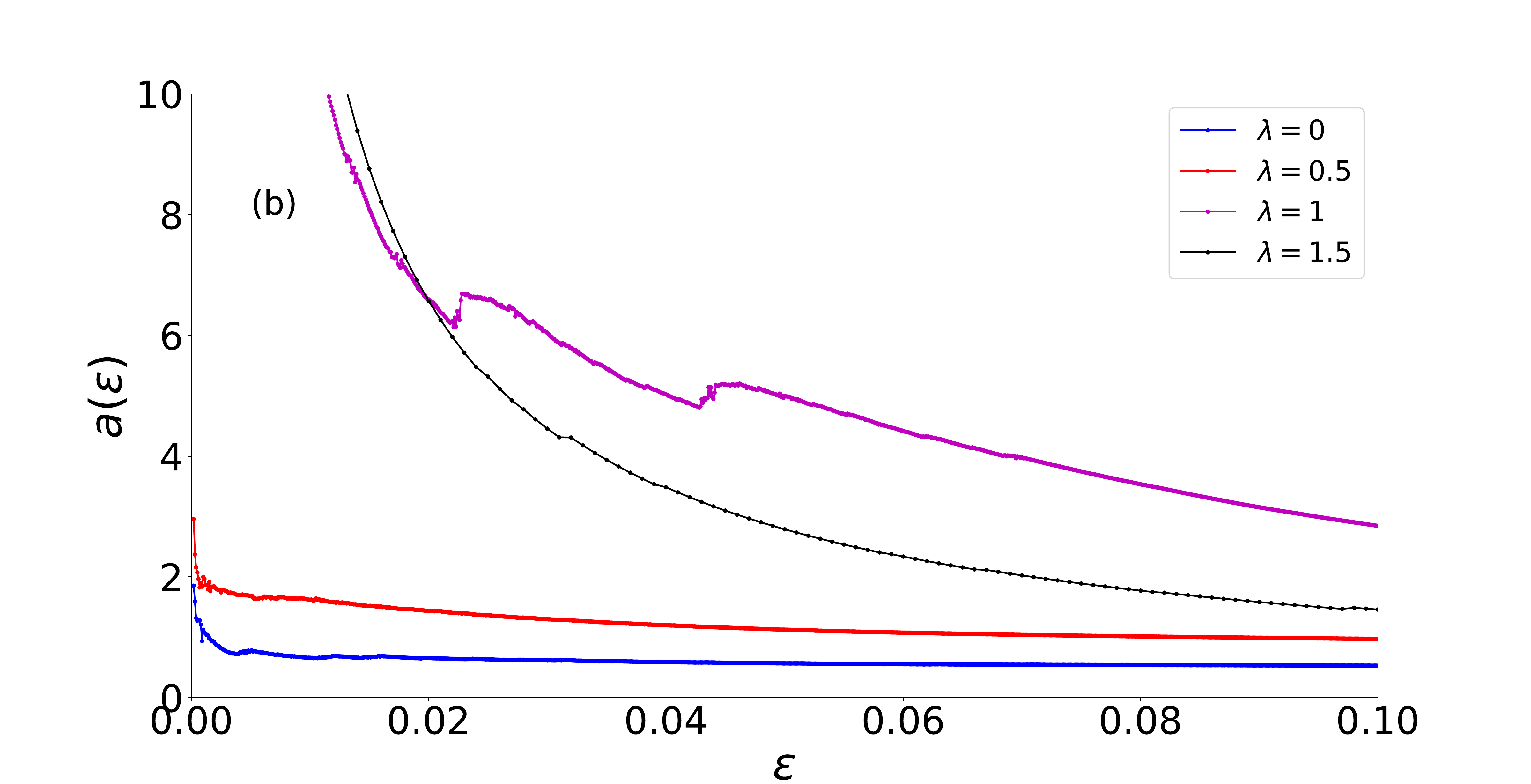}
        \caption{Numerical example of $\delta(\epsilon)$ (panel (a)) and $a(\epsilon)$ (panel (b)) at different system sizes and potential strength for $m = \frac{L}{2}, n = \frac{L}{2}+6$, for $L=800$.}
\label{fig:bound}
\end{figure}
Now all that is left is to numerically show that $\delta(\epsilon)$ is quite small.  In Fig.
\ref{fig:bound} we show our results for  $a(\epsilon)$ and $\delta(\epsilon)$
at different system sizes and potential strength.  From these results we can
conclude that the bound performs poorly in the localized regime, and at the
critical point of the model, while in the extended regime the bound appears to
perform quite well. For the extended regime it appears we may pick an
$a(\epsilon) \sim O(1)$ while picking $\delta(\epsilon) \approx 0$, meaning in these cases
we expect the OTOC to equilibrate to its infinite time average. 

\begin{figure}[t]
\centering
\includegraphics[width=\linewidth]{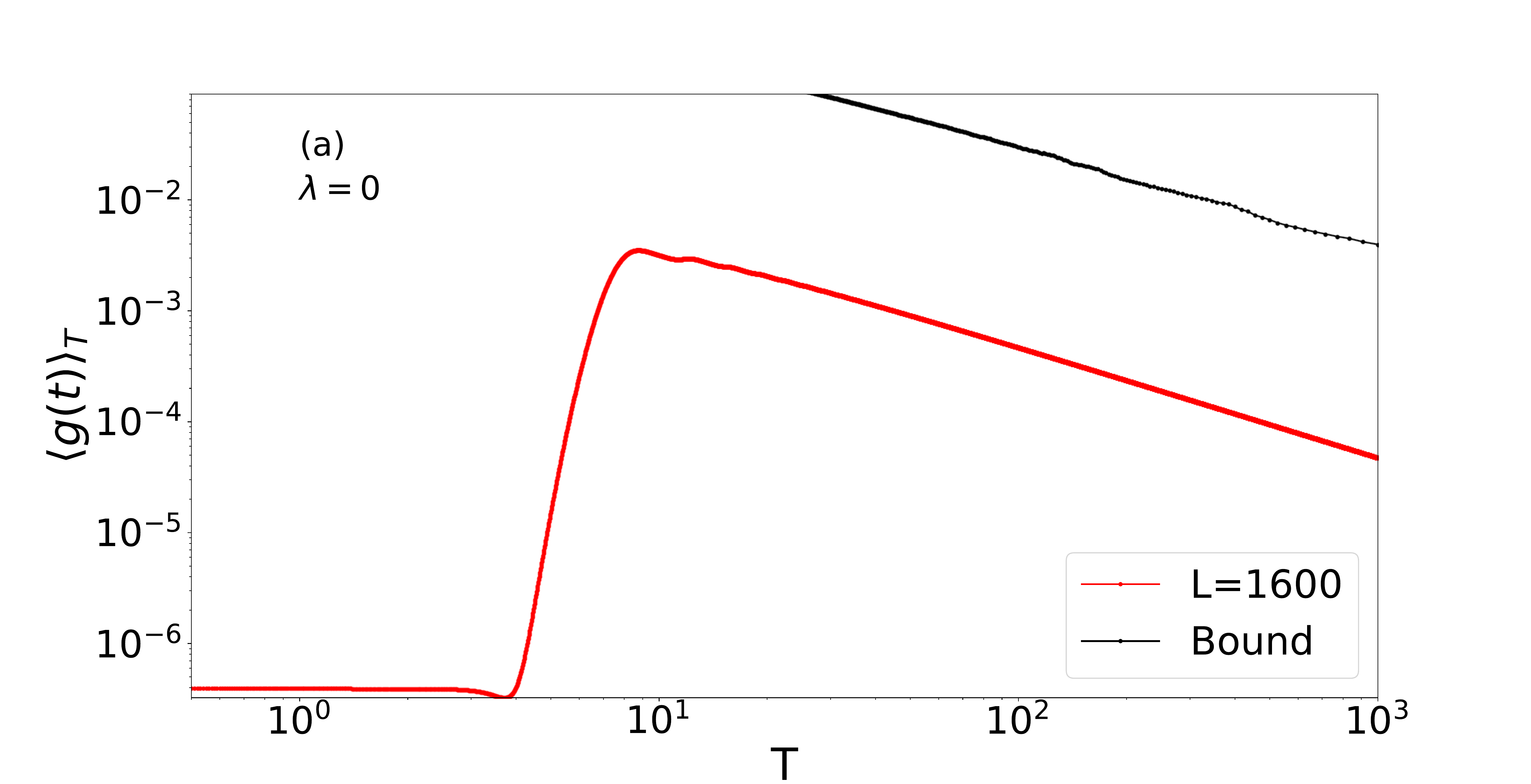}
\includegraphics[width=\linewidth]{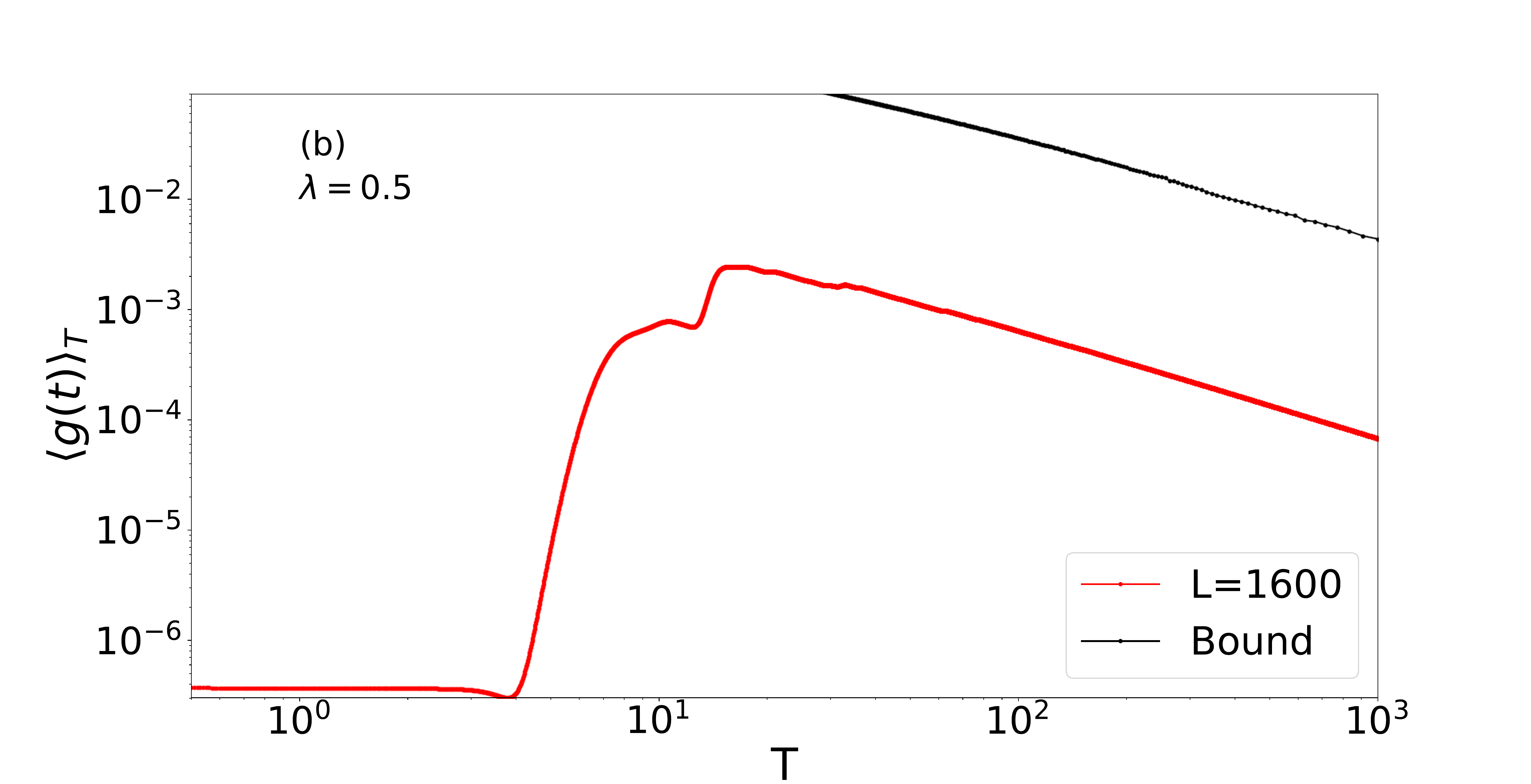}
\caption{Bound from Eq.~(\ref{eq:bound1})  for potential strengths $\lambda = 0$ (panel (a)) and $\lambda=0.5$ (panel (b)).
Both results were computed with $L=1600$ and used $m = \frac{L}{2}, n = \frac{L}{2}+6$, $L=1600$.}
\label{fig:bound1}
\end{figure}
Next we illustrate the bound, Eq.~(\ref{eq:bound1}), by numerically evaluating $\langle g_{m,n}(t) \rangle_T$ and $\xi(\frac{1}{T})$. 
Our results are shown in Fig.~\ref{fig:bound1} where we see that, as predicted, the time average defined in
Eq.~(\ref{eq:timeavg}) is not only upper bounded by Eq.~(\ref{eq:bound1}), but as the time interval $T$ is increase this upper 
bound decays to zero in the
extended region. Thus, this constitutes equilibration of an OTOC in both a
translationally invariant case ($\lambda=0$), and a case with a non-zero quasi-periodic potential ($\lambda=0.5$). This result
is expected to hold for $\lambda \in [0,\lambda_{\text{critical}})$ where for
the present numerics we have, $\lambda_{\text{critical}}=1$. Furthermore, we stress that this
result should be applicable to {\it all quadratic models} in their {\it extended} phases. 
	
Next we consider relaxation in the infinite time limit $T\to\infty$. Here, the quantity to bound (assuming for simplicity non-degenerate mode gaps, and excluding the localized and critical regimes) is, 
\begin{equation} \label{eq:infbound1}
		\lim_{T\to \infty}  \langle g_{m,n}(t) \rangle_T = \sum_{k\neq l} A_{m,k}^2A_{n,k}^2A_{m,l}^2A_{n,l}^2.
\end{equation}
From Eq.~(\ref{eq:infbound1}), using
$A_{m,k}^2 \sim \frac{1}{L}$, we see that with four such terms and only a quadratic summation over these terms 
$\lim_{T\to \infty}  \langle g_{m,n}(t) \rangle_T$ must go to zero in the extended region.
To put this into more rigorous terms we may define the constant $c = L \max_{k} \{ A_{m,k}^2,A_{n,k}^2\}$ such that, 
\begin{equation}
		\lim_{T\to \infty}  \langle g_{m,n}(t) \rangle_T \leq c^4 \sum_{k\neq l} \frac{1}{L^4} \leq \frac{c^4}{L^2},
\end{equation}
where $c$ is independent of system size due to the terms $\sqrt{L} A_{m,k} = O(1)$.

\section{Momentum OTOCs} \label{sec:momentum}

In this section we study the out of time order correlators with momentum number operators, and set, 
\begin{equation}
	\hat{A}(t) = 2\hat{\eta}_{\pi}^\dagger(t)\hat{\eta}_{\pi}(t)-1 \enspace \text{,} \enspace \hat{B} = 2\hat{\eta}_{\pi}^\dagger\hat{\eta}_{\pi}-1.
\end{equation}
The OTOC then corresponds to the $k=\pi$ momentum operator
commuting with itself in time. We make this choice since, although two momenta
$k$ and $l$ could be neighbours in momentum space, this distance isn't physical
and no wave front can be defined. The choice of $k=\pi$ is arbitrary but sits
in the "middle" of momentum space. To distinguish our results from the previous sections, where real space OTOCs were discussed,
we denote the OTOC $C_p(t)$ in this section, suppressing the $x$ dependence of $C$. The system size throughout this section is
set to $L=400$, no significant differences were observed for systems sizes up
to $L=1200$.

\subsection{Quenching}
The momentum OTOCs are studied by quenching from the ground state of the initial Hamiltonian. 
This is done in a manner identically to section \ref{Sec:posOTOC}. First we consider quenching from an initial potential strength$\lambda_i = 0$.  
\begin{figure}[h!]
\centering
\includegraphics[width=\linewidth]{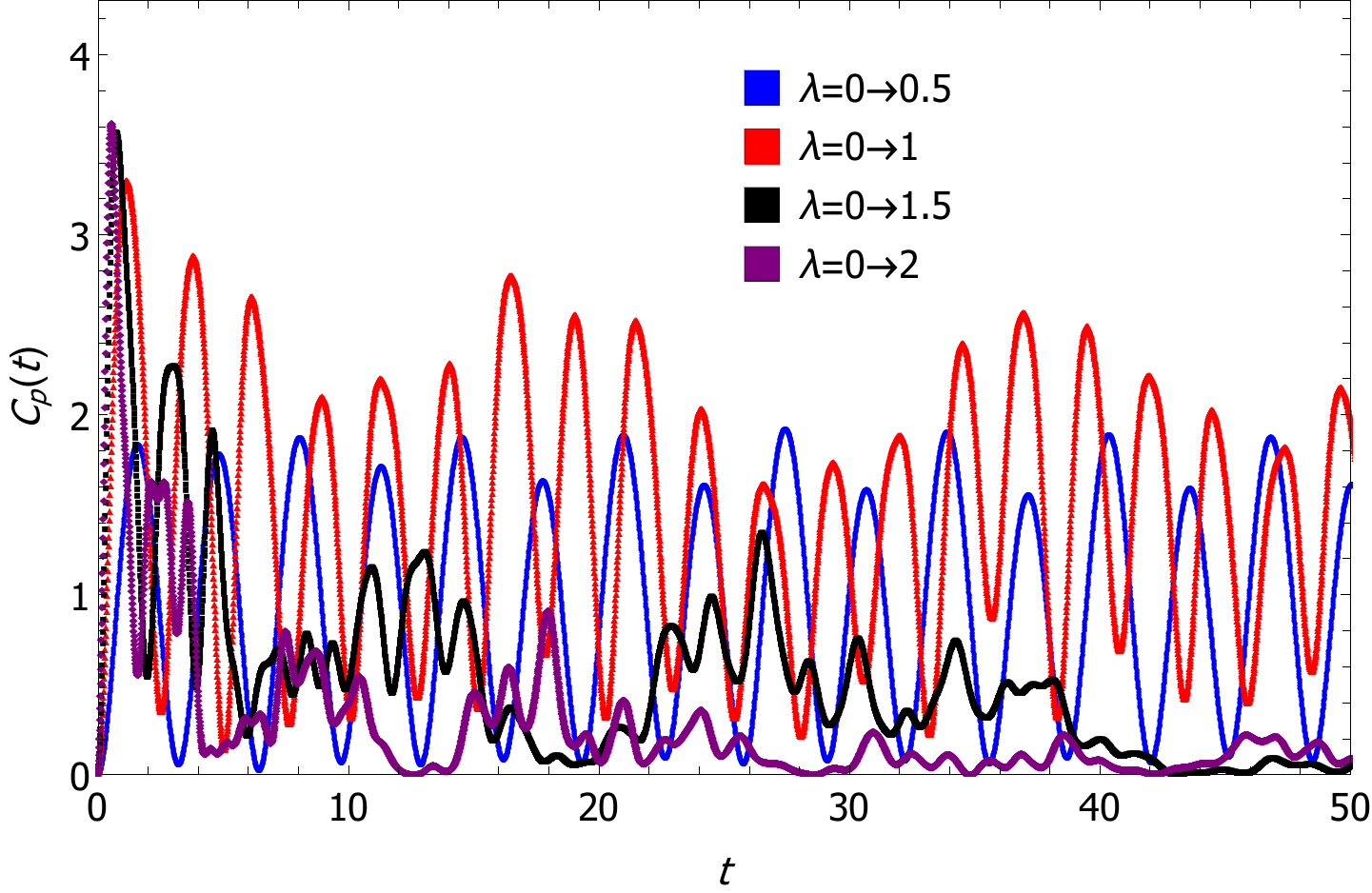}
\caption{$C_p(t)$ plotted from the dynamics of a ground state of a Hamiltonian characterized by $\lambda_i = 0$ to various final 
        Hamiltonians. This corresponds to quenching from the extended region into the critical point at $\lambda = 1$, 
        extended phase $\lambda= 0.5$ and two examples of the localized phase $\lambda = 1.5,2$. Results are for $L=400$.}
\label{fig:pqlam0midOTOCL400}
\end{figure}
	
Fig.~\ref{fig:pqlam0midOTOCL400} shows $C_p(t)$ quenched from the ground state
of the Hamiltonian with $\lambda_i = 0$ then quenched and time evolved with new
values of $\lambda_f=0.5, 1, 1.5, 2$.  Interestingly, the OTOCs all attain a
maximum, at quite early times $t<4$, and then display a slow decay from the
largest value.  The localized phase dynamics for potential strengths
of$\lambda_f = 1.5,2$ clearly show that the momentum OTOC eventually decay to
zero, and oscillate near it. The extended phase oscillates away from zero, but
does not appear to reach it.  At the critical point, $\lambda_c=1$ pronounced
oscillations is observed exceeding all other $\lambda_f$.  The extended state
is characterized by oscillations around a fixed non-zero with this value rising
with $\lambda_f$ as it approaches $\lambda_f=\lambda_c$.  

As can be clearly seen from Fig.~\ref{fig:pqlam0midOTOCL400}, the dynamics are
quite complex and it is desirable to understand the asymptotic behaviour at
the wavefront, which we can tentatively define as the first occurrence where $C(t)$
decreases. Since the momentum OTOCs are highly non-local in real space the proposed universal form,
Eq.~(\ref{eq:universalform}), is not directly applicable and we therefore consider an ad-hoc form
\begin{equation}
\label{momfit}
f(x) = C\exp(at +b/t+c/t^2+d/t^3).
\end{equation}
\begin{figure}[t]
\centering
\includegraphics[width=\linewidth]{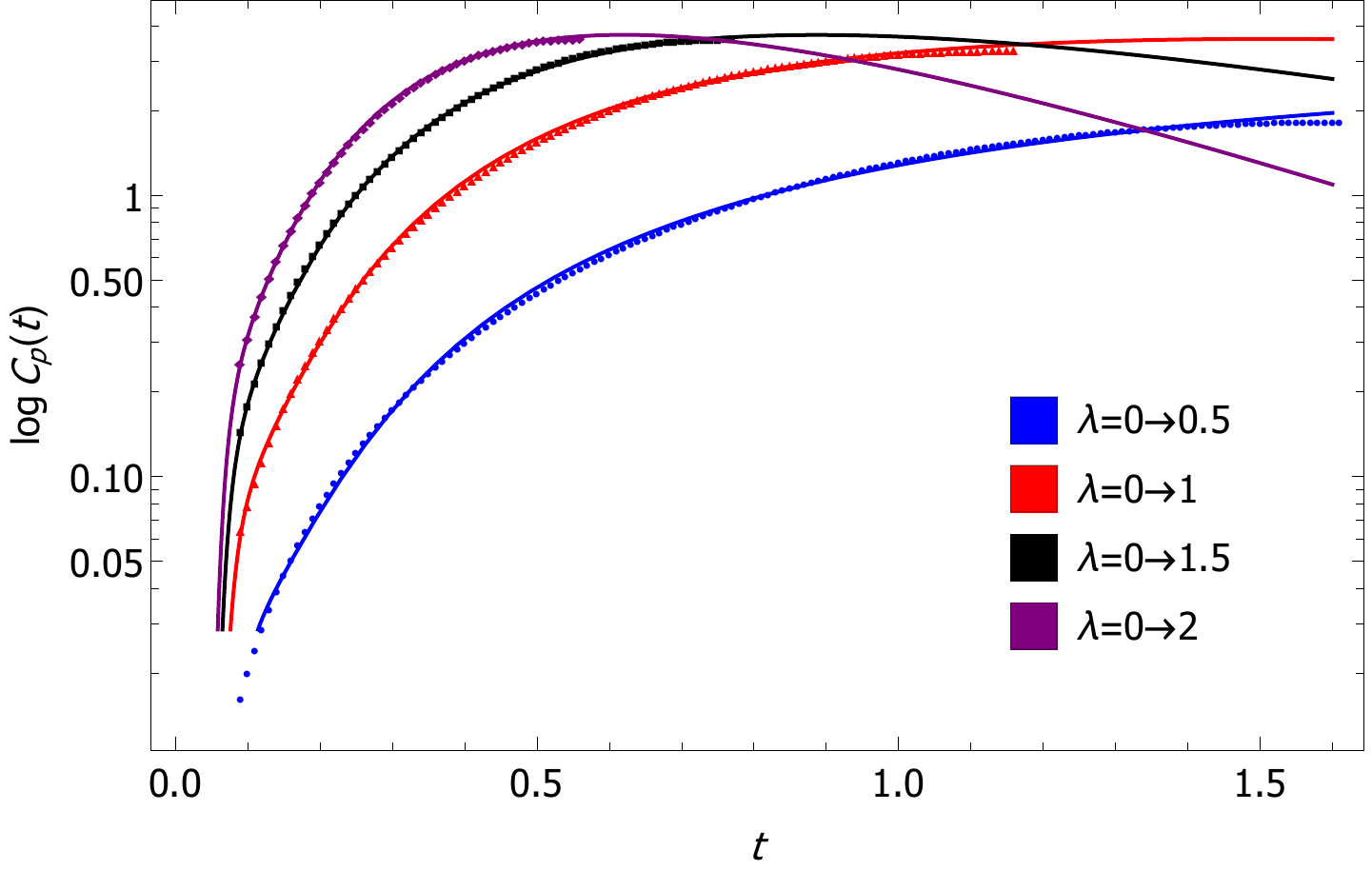}
\caption{$\log C_p(t)$ plotted from the dynamics of a ground state of a Hamiltonian characterized by $\lambda_i = 0$ to 
        various final Hamiltonians characterized by different $\lambda_f$. 
        This data is then fitted to the function, $f(t) = C\exp(at +b/t+c/t^2+d/t^3) $, which is then graphed. Results are for $L=400$}
\label{fig:pqlam0midfitL400}
\end{figure}
Results by fitting to the form, Eq.~(\ref{momfit}) are shown in 
Fig.~\ref{fig:pqlam0midfitL400} for several different values of $\lambda_f$.
Extremely good fits are obtained and we have verified that adding more terms does not significantly
improve the fits.

\begin{figure}[h!]
\centering
\includegraphics[width=\linewidth]{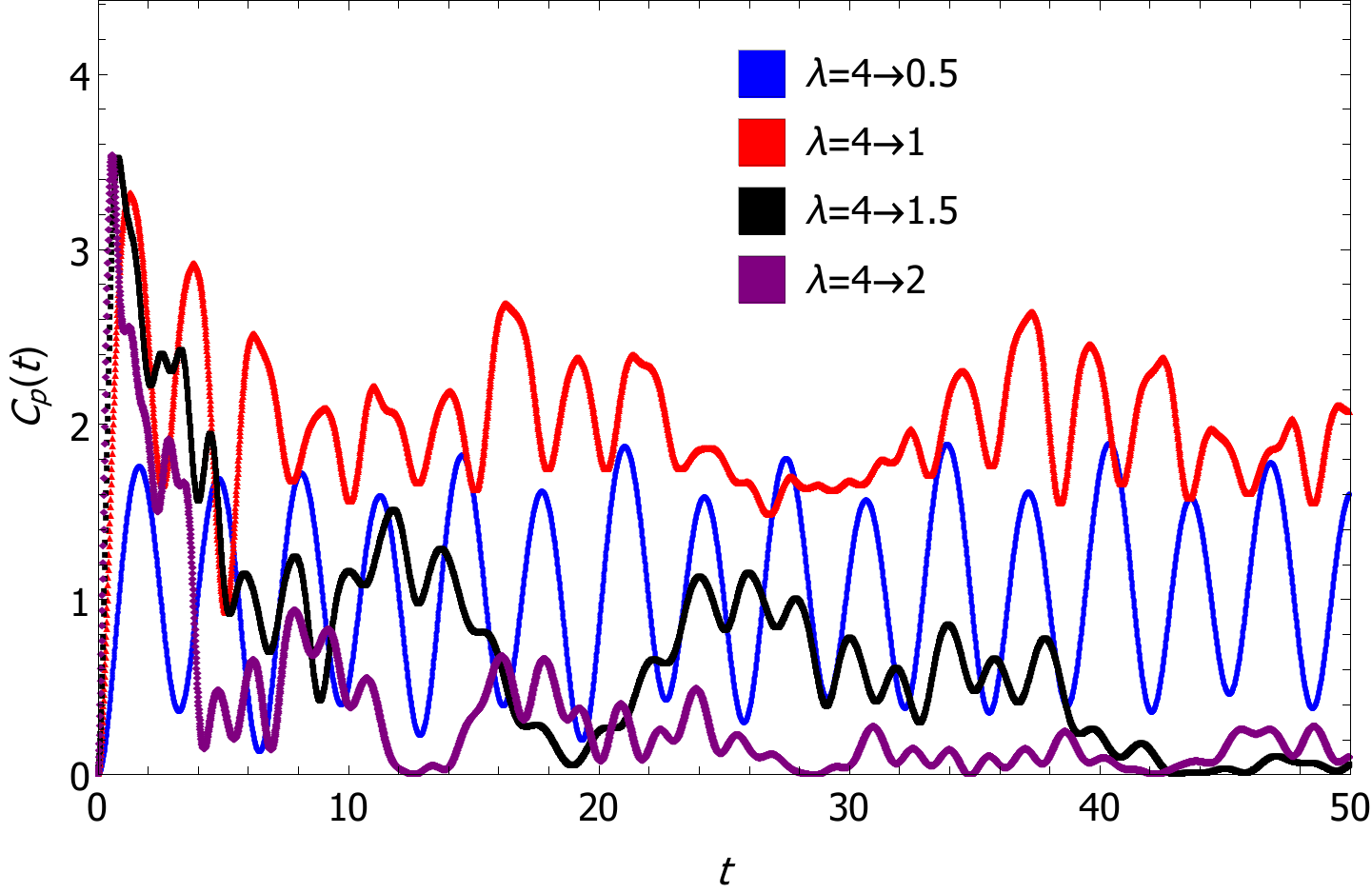}
\caption{$C_p(t)$ plotted from the dynamics of a ground state of a Hamiltonian characterized by $\lambda = 4$ 
        to various final Hamiltonians. 
        This corresponds to quenching from the extended region into the critical point at $\lambda = 1$, 
        extended phase $\lambda= 0.5$ and two examples of the localized phase $\lambda = 1.5,2$. Results are for $L=400$.}
\label{fig:pqlam40midOTOCL400}
\end{figure}
Next we consider a different quench where we instead start from the localized phase with $\lambda_i=4$ and evolve with the four 
different $\lambda_f=0.5, 1, 1.5, 2$. 
Our results for this case are shown in Fig.~\ref{fig:pqlam40midOTOCL400}. 
The oscillations in this case comparably to quecnhing from $\lambda_i=0$ shown in Fig.~\ref{fig:pqlam0midOTOCL400}.
However, their quasi-periodicity is much smaller and less chaotic. Both
examples, $\lambda_i=0,4$, are characterized by the same oscillations that appear to never
dissipate. However, the wavefront for $\lambda_i=4$ is near identical to the one shown in
Fig.~\ref{fig:pqlam0midfitL400} for $\lambda_i=0$. The same function, Eq.~(\ref{momfit}), used to fit the results for $\lambda_i=0$  can be used to 
characterize the wavefront for $\lambda_i=4$ producing extremely high quality fits almost indistinguishable from the fits 
shown in Fig.~\ref{fig:pqlam0midfitL400}.
Thus we conclude that the initial rise of the OTOC goes like Eq.~(\ref{momfit}) in
both quench scenarios. The form given in Eq.~(\ref{eq:momanti}) was also
observed to hold for momentum OTOCs defined in a thermal states, as well as for initial states in the form of a  product state:
\begin{equation} \label{prodstate} |\psi \rangle  = \prod_{l\in
\mathbb{S}} \hat{f}_l^\dagger |0 \rangle.  
\end{equation} 
where $\mathbb{S} = \{ l \in \mathbb{N}: l \mod 2 = 0 \}$. 
This then allows us to conclude that this form of the wavefront for momentum OTOCs is rather generic, and doesn't
depend on initial conditions.  

\section{Conclusion}
The AA model with a quasi-periodic potential represents a unique opportunity to investigate
quantum information dynamics in the presence of a phase transition between an extended 
localized phase using exact numerics. Here we have explicitly demonstrated equilibration of the
real-space OTOCs to {\it zero} in the extended phase of the model, a result that generalizes to any
model with quadratic interactions in an extended regime. The early time behavior of the real-space OTOCs
are largely independent of the strength of the quasi-periodic potential and follow a simple power-law with position dependent exponent
even in the localized phase. The regime close to the classical wavefront, $x=v_Bt$,  has been shown to propagate as a Gaussian (Eq. \ref{eq:Gaussian}) with distance dependent parameters which converge to constants in the large distance limit, signifying a fifth time regime of interest for the OTOC. At earlier times $t \ll \frac{x}{v_B}$ it is possible to apply the universal waveform Eq. \ref{eq:universalform} which is often applied to thermal OTOCs at infinite temperature. The spreading of information in momentum space as obtained from analyzing momentum space OTOCs is significantly
more complex and a complete understanding is currently lacking. Here we propose an ad-hoc form for the early time behaviour of the momentum OTOCs
that seem to work exceedingly well.

\section{Acknowledgements}
J.R would like to thank \'Alvaro M. Alhambra, Luis Pedro Garc\'ia-Pintos and Shenglong Xu for helpful discussions. 
This research was supported by NSERC and
enabled in part by support provided by (SHARCNET) (www.sharcnet.ca) and Compute/Calcul Canada (www.computecanada.ca).


	\onecolumngrid
\appendix

\section{Time evolution}\label{app:tevolve}
In this appendix entry we review time evolution of free fermions and present out the numerical method required to carry out of quench protocol. For more detailed treatments of the time evolution of free fermions see \cite{Riddell2},\cite{timeevolvefermions}. We are given in general a Hamiltonian written in the form, 

	\begin{equation}
	\label{locHam}
	\hat{H} = \sum_{i,j}M_{i,j}\hat{f}_{i}^{\dagger}\hat{f}_{j}.
	\end{equation}

Where we assume $M$ is real symmetric and thus can be diagonilized with a real orthogonal matrix $A$ such that $M=ADA^\dagger$. This solves the model, and we recover new fermionic operators and a diagonal Hamiltonian, 
\begin{equation}
\label{diagham}
		\hat{H} = \sum_{k}\epsilon_{k}\hat{d}_{k}^\dagger\hat{d}_{k},
\end{equation}

where we refer to $\epsilon_{k}$ as energy eigenmodes which are the entries of the diagonal matrix and the corresponding space, eigenmode space (normal modes is also regularly used). Since the states we are interested in are Gaussian (product states, thermal states, ground states), we can completely deduce all statistics of the model with the occupation matrix. Defining arbitrary fermionic operators as $\hat{b}_k^\dagger$, $\hat{b}_l$ we define the matrix in $b$ space as, 

\begin{equation}
\label{occmatrix}
	\Lambda_{k,l}^{(b)} = \langle \hat{b}_k^\dagger\hat{b}_l \rangle.
\end{equation}
Where the superscript denotes the space we are describing. In this document we refer to real space with $f$, eigenmode space with $d$ and momentum space with $p$ superscripts. 
Time evolving individual eigenmodes is easily deduced from Eq.~(\ref{diagham}), 
\begin{equation}
	\hat{d}_k(t) = e^{-i\epsilon_kt}\hat{d}_k.
\end{equation} 
For the creation operators simply take the Hermitian adjoint.  As seen in Eq.~(\ref{eq:OTOCeq}), we are interested in time evolving one or two operators in the expectation value. Thus we see that evolving the whole matrix in real space we get, 

        \begin{equation}
	\Lambda^{(f)}(t,t) = A e^{iDt} \Lambda^{(d)} e^{-iDt}A^T.
        \end{equation}
        
Where the double time arguments signify we are time evolving both the creation and annihilation part. Similarly the out of time correlations in real space can be calculated from, 

                \begin{equation}
		\Lambda^{(f)}(t,0) = A e^{iDt} \Lambda^{(d)} A^T \enspace \text{,} \enspace
		 	\Lambda^{(f)}(0,t) = A \Lambda^{(d)} e^{-iDt}A^T.
                \end{equation}
From here we can calculate the correlation functions of the momentum operators given by,
	\begin{eqnarray*}
	\eta_k := \frac{1}{\sqrt{L}} \sum_j e^{ikj} \hat{f}_j, \\
	\eta_k^\dagger := \frac{1}{\sqrt{L}} \sum_j e^{-ikj} \hat{f}_j^\dagger.
\end{eqnarray*} 
Then the correlations in momentum space are given by, 

\begin{equation}
	\Lambda^{(p)}_{k,l} = \sum_{m,n} e^{-i(mk-nl)} \Lambda_{m,n}^{(f)}.
\end{equation}
The time evolution is then found by time evolving $\Lambda_{m,n}^{(f)}$ in the desired way. Now all we need to describe is the out of time anti-commutation relations. For the real space operators, 
 \begin{eqnarray}
\label{anticom}
\{ \hat{f}_m^\dagger(t), \hat{f}_n\} 
= \sum_{k} {A}_{m,k} A_{n,k}e^{i \epsilon_k t} = a_{m,n}(t), 
         \label{eq:A8}
\end{eqnarray}
simply taking the conjugate recovers the relationship where $\hat{f}_n$ is time evolved. We also have,  $ \{ \hat{f}_m(t), \hat{f}_n\} =  \{ \hat{f}_m^\dagger(t), \hat{f}_n^\dagger\}=0$. For the momentum operators , 

	\begin{eqnarray} \label{eq:momanti}
	\label{anticomp}
	\{\eta_k^\dagger(t), \eta_p\} = \frac{1}{L} \sum_{m,n} e^{-i(km-pn)}\left(\hat{f}_m^\dagger(t) \hat{f}_n+ \hat{f}_n\hat{f}_m^\dagger(t)\right) 
	= \frac{1}{L} \sum_{m,n} e^{-i(km-pn)} a_{m,n}(t) = u_{k,p}(t).
\end{eqnarray}
Eq.~(\ref{anticomp}) is simply a discrete Fourier transform of Eq.~(\ref{anticom}). With these pieces we can now calculate the necessary correlators and out of time anti-commutators for the OTOC. 

\section{Quench protocol}\label{app:quenchprotocol}
We now turn to a discussion of the quench protocol.  We define two Hamiltonians written
identically to the one written in Eq.~(\ref{locHam}), with $\hat{H}^{(1)}$ and
$\hat{H}^{(2)}$. We first prepare the ground state of $\hat{H}^{(1)}$ by
diagonalizing $M^{(1)}$, let $\epsilon_{k}^{(1)}$ be its eigenvalues, and
preparing the eigenmode state with, 
                                \begin{equation}
\Lambda_{k,l}^{(d,1)} = \langle \hat{d}_k^\dagger \hat{d}_l \rangle =  \left\{
\begin{array}{cl}
1 & k=l \land \epsilon_{k}^{(1)}<0  \\
0 &\text{otherwise}.
\end{array}\right.
                                \end{equation}
Note that in some cases we might have $\epsilon_{k}^{(1)}=0$ for some value of $k$, making the ground state degenerate. We then choose to construct the ground state which only has negative eigenmodes occupied and neglect the zero.  We then transform the occupation matrix to real space, 
                                        \begin{equation}
	\Lambda^{(f)}(0,0) = A^{(1)T} \Lambda^{(d,1)}A^{(1)}.
                                        \end{equation}
This gives us the initial correlation functions. Next we imagine suddenly changing the Hamiltonian to $\hat{H}^{(2)}$. We can now find this states representation in the eigenmode of the new Hamiltonian by using its orthogonal transform, $\Lambda^{(d,2)} = A^{(2)T} \Lambda^{(f)} A^{(2)}$  Thus the time evolution we are interested in is written as, 
\begin{eqnarray}
\label{quenchdynamics}
	\Lambda^{(f)}(t,t) = A^{(2)} e^{iD^{(2)}t} \Lambda^{(d,2)} e^{-iD^{(2)}t}  A^{(2)T}, \\ \Lambda^{(f)}(t,0) = A^{(2)} e^{iD^{(2)}t} \Lambda^{(d,2)}  A^{(2)T}, \\
	\Lambda^{(f)}(0,t) = A^{(2)}  \Lambda^{(d,2)} e^{-iD^{(2)}t}  A^{(2)T}.
\end{eqnarray}

This representation allows us to compute statistic we could be interested in for a Gaussian state.

\section{Calculating the OTOCs}\label{app:calcotoc}
Here we present the calculation of the OTOCs in terms of second moments. In all three cases we are interested in; product states, thermal states and ground states, are Gaussian. Thus we can use Wick's theorem to calculate the OTOC. This is done similarly to \cite{Riddell2}.  Here we present the derivation for $F_b(x,t)$ for arbitrary lattice points and fermionic operators. Consider arbitrary fermionic operators $\hat{b}_i$ such that $\{\hat{b}_k,\hat{b}_l \} =\{\hat{b}_k^{\dagger},\hat{b}_l^{\dagger}\}= 0$, $\{\hat{b}_l^{\dagger},\hat{b}_k\} = \delta_{l,k}$ and $ a_{m,n}(t) = \{ \hat{b}_m^\dagger(t), \hat{b}_n\} $, where we assume $ \{ \hat{b}_m(t), \hat{b}_n\} =  \{ \hat{b}_m^\dagger(t), \hat{b}_n^\dagger\} = 0$. Then we are interested in the real part of the function,

                                                \begin{equation}
	F_b(x,t) = \langle \left( \hat{b}_i^\dagger(t) \hat{b}_i(t) -\frac{1}{2}     \right)  \left( \hat{b}_j^\dagger \hat{b}_j -\frac{1}{2}     \right) \left( \hat{b}_i^\dagger(t) \hat{b}_i(t) -\frac{1}{2}     \right)  \left( \hat{b}_j^\dagger \hat{b}_j -\frac{1}{2}     \right) \rangle. 
                                                \end{equation}

Adopting the notation $\hat{n}_i =  \hat{b}_i^\dagger \hat{b}_i$ and using  $\hat{n}_i(t)^2 = n_i(t)$ we can write,

\begin{eqnarray}
F(t) =16 \langle \hat{n}_{i}(t) \hat{n}_{j} \hat{n}_{i}(t) \hat{n}_{j}  -\frac{1}{2}(\hat{n}_{i}(t) \hat{n}_{j} \hat{n}_{i}(t) +   \hat{n}_{j} \hat{n}_{i}(t) \hat{n}_{j} ) + \frac{1}{4}(\hat{n}_{j} \hat{n}_{i}(t) - \hat{n}_{i}(t) \hat{n}_{j} ) + \frac{1}{16} \rangle. 
\end{eqnarray}

Here we present the derivation for the thermal state, but since all states considered are Gaussian the end result will be equivalent. 
Throughout the derivation we abuse the fact that $\hat{b}_i^2 = \left(\hat{b}_i^\dagger\right)^2 = 0 $, the out of time anti-commutation rules, and assuming that each $\hat{b}_k$ is a linear combination of $\hat{d}_l$ terms only. 
Now we can focus on treating each term based on our initial conditions as before. 
Let us deal with each term of $F(t)$ individually. First consider the fourth order correlations,

                                                        \begin{equation}
\langle \hat{n}_{j} \hat{n}_{i}(t) - \hat{n}_{i}(t) \hat{n}_{j} \rangle_\beta.
                                                        \end{equation}
Let us derive a rule to contract  these fourth moments. Consider, 

\begin{eqnarray}
	\langle 	\hat{n}_{i}(t) \hat{n}_{j} \rangle_\beta = \langle \hat{b}_i^\dagger(t) \hat{b}_i(t) \hat{b}_j^\dagger \hat{b}_j \rangle_\beta = \sum_{m,n,k,l} A_{i,k} A_{i,l}A_{j,m}A_{j,n} e^{i(\epsilon_{k}-\epsilon_{l}t)} \langle \hat{d}_k^\dagger \hat{d}_l \hat{d}_m^\dagger \hat{d}_n \rangle_\beta .
\end{eqnarray}

Using the fact that,

\begin{equation}
\label{thermOcc}
\Lambda_{k,l}^{d,\beta} = \langle \hat{d}_k^\dagger \hat{d}_l \rangle_\beta =  \left\{
\begin{array}{cl}
\frac{1}{1+e^{\beta \epsilon_k}} & k=l,  \\
0 &\text{otherwise}.
\end{array}\right.
\end{equation}

\begin{eqnarray}
	\tr \left( \hat{d}_k^\dagger \hat{d}_l \hat{d}_m^\dagger \hat{d}_n \rho_\beta \right) = \delta_{k,l} \tr(\hat{d}_m^\dagger \hat{d}_n \rho_\beta)+\delta_{k,n} \tr (\hat{d}_l\hat{d}_m^\dagger \rho_\beta ) - \tr (  \hat{d}_l \hat{d}_m^\dagger \hat{d}_n \hat{d}_k^\dagger \rho_\beta  ).
\end{eqnarray}
	Using $e^{-\beta \epsilon_{k} \hat{n}_k} \hat{d}_k^\dagger = e^{-\beta \epsilon_{k}} \hat{d}_k^\dagger e^{-\beta \epsilon_{k} \hat{n}_k}$ we get, 
\begin{eqnarray}
	(1+e^{\beta \epsilon_{k}})\tr \left( \hat{d}_k^\dagger \hat{d}_l \hat{d}_m^\dagger \hat{d}_n \rho_\beta \right) = \delta_{k,l} \tr(\hat{d}_m^\dagger \hat{d}_n \rho_\beta)+\delta_{k,n} \tr (\hat{d}_l\hat{d}_m^\dagger \rho_\beta ), \\
	\implies \tr \left( \hat{d}_k^\dagger \hat{d}_l \hat{d}_m^\dagger \hat{d}_n \rho_\beta \right) =   \langle \hat{d}_k^\dagger \hat{d}_{l} \rangle   \tr(\hat{d}_m^\dagger \hat{d}_n \rho_\beta)+ \langle \hat{d}_k^\dagger \hat{d}_{n} \rangle  \tr (\hat{d}_l\hat{d}_m^\dagger \rho_\beta ).
\end{eqnarray}

This then gives, 

\begin{eqnarray}
	\langle 	\hat{n}_{i}(t) \hat{n}_{j} \rangle_\beta 
	=  \langle \hat{b}_i^\dagger(t) \hat{b}_{i}(t) \rangle_\beta   \langle\hat{b}_j^\dagger \hat{b}_j \rangle_\beta + \langle \hat{b}_i^\dagger(t) \hat{b}_{j} \rangle_\beta  \langle \hat{b}_i(t)\hat{b}_j^\dagger \rangle_\beta .
\end{eqnarray}

Similarly,
\begin{eqnarray}
	\langle  \hat{n}_{j} \hat{n}_{i}(t) \rangle_\beta =    \langle \hat{b}_j^\dagger \hat{b}_{j} \rangle_\beta   \langle\hat{b}_i^\dagger(t) \hat{b}_i(t) \rangle_\beta + \langle \hat{b}_j^\dagger \hat{b}_{i}(t) \rangle_\beta  \langle \hat{b}_j\hat{b}_i^\dagger(t) \rangle_\beta      
\end{eqnarray}
From here we see that,
\begin{eqnarray}
	&\langle  \hat{n}_{j} \hat{n}_{i}(t) \rangle_\beta - \langle \hat{n}_{i}(t) \hat{n}_{j} \rangle_\beta 
	= \langle \hat{b}_j^\dagger \hat{b}_{i}(t) \rangle_\beta  \langle \hat{b}_j\hat{b}_i^\dagger(t) \rangle_\beta -\langle \hat{b}_i^\dagger(t) \hat{b}_{j} \rangle  \langle \hat{b}_i(t)\hat{b}_j^\dagger \rangle_\beta, \\  &= \langle \hat{b}_j^\dagger \hat{b}_{i}(t) \rangle_\beta \left( a_{i,j}(t) - \langle \hat{b}_i^\dagger(t) \hat{b}_j\rangle_\beta    \right) -  \langle \hat{b}_i^\dagger(t) \hat{b}_{j} \rangle_\beta \left( \bar{a}_{i,j}(t) - \langle \hat{b}_j^\dagger \hat{b}_i(t)\rangle_\beta   \right) =
	a_{i,j}(t)\langle \hat{b}_j^\dagger \hat{b}_{i}(t) \rangle_\beta - \bar{a}_{i,j}(t)\langle \hat{b}_i^\dagger(t) \hat{b}_{j} \rangle_\beta,
\end{eqnarray} 
this is however a purely imaginary number and therefore does not contribute to the OTOC. Now the sixth order term, 
\begin{eqnarray}
	&\hat{n}_{j} \hat{n}_{i}(t) \hat{n}_{j} = \hat{b}_j^\dagger \hat{b}_j \hat{b}_i^\dagger(t)\hat{b}_i(t) \hat{b}_j^\dagger \hat{b}_j, \\
	&=\hat{b}_j^\dagger \left(a_{i,j}(t)-\hat{b}_i^\dagger(t) \hat{b}_j\right) \hat{b}_i(t) \hat{b}_j^\dagger \hat{b}_j, \\
	&= a_{i,j}(t)\hat{b}_j^\dagger \hat{b}_i(t) \hat{b}_j^\dagger \hat{b}_j - \hat{b}_j^\dagger \hat{b}_i^\dagger(t)\hat{b}_j\hat{b}_i(t) \hat{b}_j^\dagger \hat{b}_j,\\
	&= a_{i,j}\hat{b}_j^\dagger \left( \bar{a}_{i,j}(t) - \hat{b}_j^\dagger \hat{b}_i(t) \right) \hat{b}_j + \hat{b}_j^\dagger \hat{b}_i^\dagger(t) \hat{b}_i(t)\hat{b}_j \hat{b}_j^\dagger \hat{b}_j, \\
	&|a_{i,j}|^2 \hat{b}_j^\dagger \hat{b}_j + \hat{b}_j^\dagger \hat{b}_i^\dagger(t) \hat{b}_i(t)  \left( 1 - \hat{b}_j^\dagger \hat{b}_j\right) \hat{b}_j, \\
	&= |a_{i,j}|^2 \hat{b}_j^\dagger \hat{b}_j + \hat{b}_j^\dagger \hat{b}_i^\dagger(t) \hat{b}_i(t) \hat{b}_j.
\end{eqnarray}
Then applying the expectation value, 
\begin{eqnarray}
	&|a_{i,j}|^2 \langle \hat{b}_i^\dagger(t)\hat{b}_i(t) \rangle_\beta + \langle \hat{b}_i^\dagger(t)\hat{b}_j^\dagger \hat{b}_j\hat{b}_i(t) \rangle_\beta,  \\
	&= 	|a_{i,j}|^2 \langle \hat{b}_i^\dagger(t)\hat{b}_i(t) \rangle_\beta 	+ \sum_{m,n,k,l} A_{i,k} A_{j,l}A_{j,m}A_{i,n} e^{i(\epsilon_{k}-\epsilon_{n}t)} \langle \hat{d}_k^\dagger \hat{d}_l^\dagger \hat{d}_m \hat{d}_n \rangle_\beta,	\\
	&= |a_{i,j}|^2 \langle \hat{b}_i^\dagger(t)\hat{b}_i(t) \rangle_\beta 	+ \sum_{m,n,k,l} A_{i,k} A_{j,l}A_{j,m}A_{i,n} e^{i(\epsilon_{k}-\epsilon_{n}t)} \left(  -\langle \hat{d}_k^\dagger \hat{d}_m \rangle_\beta  \langle \hat{d}_l^\dagger \hat{d}_n \rangle_\beta +   \langle \hat{d}_k^\dagger \hat{d}_n \rangle_\beta \langle \hat{d}_l^\dagger \hat{d}_m \rangle_\beta  \right), \\
	&= |a_{i,j}|^2 \langle \hat{b}_i^\dagger(t)\hat{b}_i(t) \rangle_\beta + \langle  \hat{b}_j^\dagger \hat{b}_j \rangle_\beta \langle \hat{b}_i(t)^\dagger \hat{b}_i(t) \rangle_\beta  - \langle \hat{b}_j^\dagger \hat{b}_i(t) \rangle_\beta \langle \hat{b}_i^\dagger(t) \hat{b}_j \rangle_\beta 
\end{eqnarray}
Next we look at the other 6th moment, 
\begin{equation}
	\hat{n}_{i}(t) \hat{n}_{j} \hat{n}_{i}(t) =   \hat{b}_i^\dagger(t)\hat{b}_i(t) \hat{b}_j^\dagger \hat{b}_j\hat{b}_i^\dagger(t)\hat{b}_i(t).
\end{equation}
The strategy here is identical, and we arrive at, 
\begin{equation}
\hat{n}_{i}(t) \hat{n}_{j} \hat{n}_{i}(t) = |a_{i,j}|^2 \hat{b}_i^\dagger(t)\hat{b}_i(t) + \hat{b}_i^\dagger(t)\hat{b}_j^\dagger \hat{b}_j\hat{b}_i(t).
\end{equation}
Applying the thermal expectation value, 
\begin{eqnarray}
	&\langle \hat{n}_{i}(t) \hat{n}_{j} \hat{n}_{i}(t) \rangle_\beta =|a_{i,j}|^2 \langle \hat{b}_i^\dagger(t)\hat{b}_i(t) \rangle_\beta + \langle \hat{b}_i^\dagger(t)\hat{b}_j^\dagger \hat{b}_j\hat{b}_i(t) \rangle_\beta, \\
	&= |a_{i,j}|^2 \langle \hat{b}_i^\dagger(t)\hat{b}_i(t) \rangle_\beta + \langle  \hat{b}_j^\dagger \hat{b}_j \rangle_\beta \langle \hat{b}_i(t)^\dagger \hat{b}_i(t) \rangle_\beta  - \langle \hat{b}_j^\dagger \hat{b}_i(t) \rangle_\beta \langle \hat{b}_i^\dagger(t) \hat{b}_j \rangle_\beta .
\end{eqnarray}
	So we finally need the eighth order term which is made easier by knowing the results from the 6th order terms,
\begin{eqnarray}
	&\hat{n}_{i}(t) \hat{n}_{j} \hat{n}_{i}(t) \hat{n}_j =   \hat{b}_i^\dagger(t)\hat{b}_i(t) \left( \hat{b}_j^\dagger \hat{b}_j\hat{b}_i^\dagger(t)\hat{b}_i(t)\hat{b}_j^\dagger \hat{b}_j \right), \\
	&= \hat{b}_i^\dagger(t)\hat{b}_i(t) \left( |a_{i,j}(t)|^2 \hat{b}_j^\dagger \hat{b}_j + \hat{b}_j^\dagger \hat{b}_i^\dagger(t) \hat{b}_i(t) \hat{b}_j \right), \\
	&=|a_{i,j}(t)|^2 \hat{b}_i^\dagger(t)\hat{b}_i(t)\hat{b}_j^\dagger \hat{b}_j+ \hat{b}_i^\dagger(t)\hat{b}_i(t) \hat{b}_j^\dagger \hat{b}_i^\dagger(t) \hat{b}_i(t) \hat{b}_j, \\
	&= |a_{i,j}(t)|^2 \hat{b}_i^\dagger(t)\hat{b}_i(t)\hat{b}_j^\dagger \hat{b}_j + \hat{b}_j^\dagger \hat{b}_i^\dagger(t) \hat{b}_i(t) \hat{b}_j.
\end{eqnarray}
Now, taking the thermal expectation value we can use previous results, (the first term is from the fourth moments, and second from the sixth) , 
\begin{eqnarray} 
	&=	|a_{i,j}(t)|^2 ( \langle \hat{b}_i^\dagger(t) \hat{b}_i(t)\rangle_\beta  \langle \hat{b}_j^\dagger \hat{b}_j\rangle_\beta +  \langle \hat{b}_i^\dagger(t) \hat{b}_j\rangle_\beta  \langle   \hat{b}_i(t)\hat{b}_j^\dagger \rangle_\beta )  
	+  \langle \hat{b}_i^\dagger(t) \hat{b}_i(t)\rangle_\beta  \langle \hat{b}_j^\dagger \hat{b}_j\rangle_\beta -  \langle \hat{b}_i^\dagger(t) \hat{b}_j\rangle_\beta  \langle \hat{b}_j^\dagger \hat{b}_i(t)\rangle_\beta,   
	\\
	&= |a_{i,j}(t)|^2 ( \langle \hat{b}_i^\dagger(t) \hat{b}_i(t)\rangle_\beta  \langle \hat{b}_j^\dagger \hat{b}_j\rangle_\beta +  \langle \hat{b}_i^\dagger(t) \hat{b}_j\rangle_\beta  \langle \bar{a}_{i,j}(t)- \hat{b}_j^\dagger \hat{b}_i(t)\rangle_\beta )  
	+  \langle \hat{b}_i^\dagger(t) \hat{b}_i(t)\rangle_\beta  \langle \hat{b}_j^\dagger \hat{b}_j\rangle_\beta -  \langle \hat{b}_i^\dagger(t) \hat{b}_j\rangle_\beta  \langle \hat{b}_j^\dagger \hat{b}_i(t)\rangle_\beta   
\end{eqnarray}

Grouping everything together finally gives us,
\begin{eqnarray}
\label{eq:OTOCeq}
F_b(x,t) = 16|a_{i,j}(t)|^2 \left( \langle \hat{b}_i^\dagger(t) \hat{b}_i(t)\rangle_\beta  \langle \hat{b}_j^\dagger \hat{b}_j\rangle_\beta  
-\frac{1}{2}\left(\langle \hat{b}_i^\dagger(t) \hat{b}_i(t)\rangle_\beta+  \langle \hat{b}_j^\dagger \hat{b}_j\rangle_\beta \right) +  \bar{a}_{i,j}(t)\langle \hat{b}_i^\dagger(t)\hat{b}_j\rangle_\beta-\langle \hat{b}_i^\dagger(t)\hat{b}_j\rangle_\beta \langle \hat{b}_j^\dagger\hat{b}_i(t)\rangle_\beta \right) +1.
\end{eqnarray}
Note in the case of product states this form is significantly reduced and in the case of the ground state, one can simply drop the thermal expectation values. This form is general and recovers both cases used in \cite{Riddell2}.

\section{Bounding uniform average}\label{app:bound}
Here we provide the proof to bound the uniform average found in Eq. \ref{eq:timeavg}. This proof is similar to \cite{Malabarba14,Garcia-PintosPRX2017,Riddell3} and is provided here for completeness. 
Consider the Gaussian probability density function with average $\mu = \frac{T}{2}$ and standard deviation $\sigma = \alpha T$,

\begin{equation}
p_G(t)  = 	\frac{1}{\sqrt{2\pi \alpha^2T^2}} e^{-\frac{(t-T/2)^2}{2\alpha^2T^2}}, \enspace t\in R
\end{equation}

Similarly we define the uniform probability density function as,

\begin{equation}
\label{eq:uniform}
p_T(t) =  \left\{
\begin{array}{cl}
\frac{1}{T} & t\in [0,T],  \\
0 &\text{otherwise}.
\end{array}\right.
\end{equation}

Let $f(t)$ be some positive function of time, then the Gaussian and uniform averages are written, 

\begin{eqnarray}
	\langle f(t) \rangle_{G_T} = \int_{-\infty}^\infty f(t)p_G(t) dt, \enspace \text{and}, \enspace \langle f(t) \rangle_{T} = \int_{-\infty}^\infty f(t)p_T(t) dt.
\end{eqnarray}

We wish to find some constant $\gamma$ such that for all $t\in[0,T]$, 

\begin{equation}
	\langle f(t) \rangle_{T} \leq \gamma \langle f(t) \rangle_{G_T}.
\end{equation}
This can be made tight by setting the two probability densities identical to each other at $t=T$ and ensuring the Gaussian is larger than the uniform distribution on this interval. For the Gaussian this gives,
\begin{equation}
p_G(t=T) = \frac{1}{\sqrt{2 \pi} \alpha T} e^{-\frac{1}{8\alpha^2}},
\end{equation}
Meaning we can write, 

\begin{equation}
\langle f(t) \rangle_T \leq \gamma \langle f(t) \rangle_{G_T},
\end{equation}
where $\gamma = \gamma(\alpha) = \sqrt{2 \pi }\alpha e^{\frac{1}{8\alpha^2}}$ and $\alpha>0$ is a free parameter we can choose to minimize the constant. Next we introduce our unitary dynamics to proceed bounding the function,

\begin{equation}
	f(t) = \sum_{\alpha,\beta}p_\alpha p_\beta e^{i\left(G_{\alpha}- G_{\beta} \right)t},
\end{equation}
where $p_\alpha$ is a discrete probability distribution such that $p_\alpha \geq 0 $ and  $\sum_\alpha p_\alpha = 1$. Then we may write,

	\begin{equation}
\langle f(t) \rangle_T = \frac{1}{T} \int_{0}^T \sum_{\alpha, \beta}p_{\alpha} p_{\beta} e^{i\left(G_{\alpha}- G_{\beta} \right)t}dt \leq   \frac{\gamma}{\sqrt{2\pi \alpha^2T^2}} \int_{-\infty}^\infty \sum_{\alpha, \beta}p_{\alpha} p_{\beta} e^{i\left(G_{\alpha}- G_{\beta} \right)t} e^{-\frac{(t-T/2)^2}{2\alpha^2T^2}} dt.
\end{equation}
Let $\Delta G = G_{\alpha}- G_{\beta}$. Then each term in the sum is simply the characteristic function of the Gaussian. Using the well known identity, 

\begin{equation} \label{eq:cf}
 \frac{1}{\sqrt{2\pi \alpha^2T^2}}\int_{-\infty}^\infty e^{i\Delta G t} e^{-\frac{(t-T/2)^2}{2\alpha^2T^2}} dt = e^{i\mu \Delta G - \frac{\sigma^2 \Delta G^2}{2}}.
\end{equation}

Taking the magnitude of Eq. \ref{eq:cf} we can put everything together and write, 

\begin{equation}
	\langle f(t) \rangle_T \leq \gamma \sum_{\alpha\beta} p_\alpha p_\beta e^{ - \frac{\sigma^2 \Delta G^2}{2}}
\end{equation}
Next we introduce the function, 

\begin{equation}
g(x) = \left\{
\begin{array}{cl}
1 & \text{ if } x\in[0,1),  \\
0 &\text{otherwise}.
\end{array}\right.
\end{equation}

We also need the bound,

\begin{equation}
e^{-x^2} \leq \sum_{n=0}^{\infty} e^{-n^2} g(|x|-n)
\end{equation}
we re-express this as, 

\begin{equation}
e^{-x^2} = e^{\left(\frac{\alpha}{2}\right)^{-\Delta G^2 T^2}} \leq \sum_{n=0}^{\infty} r^{n^2} g(\Delta G^2 T^2-n), 
\end{equation}
where we must restrict ourselves to the case that $e^{\frac{\alpha}{2}} > 1$ and where $r=e^{-\frac{\alpha}{2}}$. Then,

\begin{equation}
		\langle f(t) \rangle_T \leq \gamma \sum_{n=0}^{\infty} r^{n^2} \sum_{\alpha,\beta} p_\alpha p_\beta g\left(\Delta G^2 T^2-n\right),
\end{equation}
To further break this sum up we may restrict the values of $\beta$ based on the definition the values of $\alpha$. Consider $\Delta G^2 T^2-n \in [0,1)$.

\begin{equation}
	\Delta G^2 T^2-n \in [0,1) \implies G_\beta \in I_+ =  \left[ G_\alpha+ \frac{\sqrt{n}}{T}, G_\alpha+\frac{\sqrt{n+1}}{T} \right) \enspace \text{and} \enspace I_- = \left( G_\alpha -  \frac{\sqrt{n+1}}{T}, G_\alpha+\frac{\sqrt{n}}{T} \right].
\end{equation}
The length of this interval is upper bounded by $\sqrt{n+1} - \sqrt{n} \leq 1$, $n\in \mathbb{N} \cup\{0\}$. Thus we can introduce the function, 

\begin{equation}
\xi_p(x) = \max_{\beta} \sum_{\alpha:G_\beta \in [G_\beta, G_\beta+x]} p_\alpha,
\end{equation}

which allows us to finally write, 

\begin{equation}
\langle f(t) \rangle_T \leq \gamma \sum_{n=0}^{\infty} r^{n^2} \sum_{\alpha}p_\alpha \left( \sum_{G_\beta \in I_+} p_\beta + \sum_{G_\beta \in I_-} p_\beta \right)  \leq 2 \gamma \xi_p\left(\frac{1}{T}\right) \sum_{n=0}^\infty r^{n^2},
\end{equation}

then it remains to minimize the constant term. The sum is related to the elliptic theta function by $\sum_{n=0}^\infty r^{n^2} = \frac{1}{2} \left( \Theta_3(0,r)+1  \right)$ which is convergent for all $r<1$. The entire constant is minimized by $\alpha \approx 0.6347$, which gives, 

\begin{equation}
	\langle f(t) \rangle_T \leq \kappa \pi \xi_p\left(\frac{1}{T}\right)
\end{equation} 

where $\kappa \approx 2.8637$. This completes the proof.

\twocolumngrid
\bibliography{references}

\end{document}